\documentclass[reprint,%
 amssymb, amsmath,%
 prc,cha,%
superscriptaddress
]{revtex4-2}
\usepackage[colorlinks=true,linkcolor=blue]{hyperref}%
\usepackage{physics}
\usepackage{graphicx}
\usepackage{lipsum}
\usepackage{mwe}
\usepackage{caption}
\usepackage{booktabs}
\usepackage{color}
\usepackage{soul}

\begin{document}

\title{Nuclear mass predictions using machine learning models}%

\author{Esra Y\"uksel}
\email{e.yuksel@surrey.ac.uk}
\affiliation{Department of Physics, University of Surrey, Guildford, Surrey GU2 7XH, United Kingdom}

\author{Derya Soydaner}
\email{derya.soydaner@kuleuven.be}
\affiliation{Department of Brain and Cognition, University of Leuven (KU Leuven), Leuven, Belgium}

\author{H\"useyin Bahtiyar}
\email{huseyinbahtiyar@gmail.com}
 \affiliation{Y. Tahsinbey St. 8/4, 34841, Maltepe, Istanbul, T\"urkiye}

\begin{abstract}
The exploration of nuclear mass or binding energy, a fundamental property of atomic nuclei, remains at the forefront of nuclear physics research due to limitations in experimental studies and uncertainties in model calculations, particularly when moving away from the stability line. In this work, we employ two machine learning (ML) models,  Support Vector Regression (SVR) and Gaussian Process Regression (GPR), to assess their performance in predicting nuclear mass excesses using available experimental data and a physics-based feature space. We also examine the extrapolation capabilities of these models using newly measured nuclei from AME2020 and by extending our calculations beyond the training and test set regions. Our results indicate that both SVR and GPR models perform quite well within the training and test regions when informed with a physics-based feature space. Furthermore, these ML models demonstrate the ability to make reasonable predictions away from the available experimental data, offering results comparable to the model calculations. Through further refinement, these models can be used as reliable and efficient ML tools for studying nuclear properties in the future.

\end{abstract}

\date{\today}%

\maketitle

\section{Introduction}
The atomic nucleus, a strongly correlated many-body system, is characterized by its proton ($Z$) and neutron ($N$) numbers. Mass is a fundamental property of atomic nuclei, playing a crucial role in our understanding of strong nuclear interactions and a vital role in nuclear astrophysical calculations, such as r-process simulations, where they serve as inputs \cite{PhysRevC.92.035807,PhysRevLett.116.121101}. Experimentally, it is possible to study nuclei near the stability lines, and accurate nuclear data for the masses of nuclei are available \cite{Wang_2017,Wang_2021}. However, despite significant advancements in nuclear facilities, measurements on the neutron-rich side of the nuclear chart remain unfeasible in the near future, and the exploration of the majority of nuclei involved in the r-process has yet to be undertaken. For instance, the neutron drip line, which indicates the position of the last bound nucleus, has been confirmed only up to Z = 10 \cite{PhysRevLett.123.212501}, and the boundaries of the nuclear landscape are not known experimentally. Therefore, our understanding of nuclear properties away from the stability line and the limits of the nuclear landscape relies heavily on theoretical calculations.

Up to now, several theoretical models have been used to investigate nuclear properties and determine the location of drip lines. Within this framework, microscopic-macroscopic (mic-mac) global nuclear mass models, such as Weizs\"{a}cker-Skyrme Nuclear Mass Tables (WS4) \cite{WANG2014215} and Finite-Range Droplet Model (FRDM-2012) \cite{MOLLER20161}, have demonstrated considerable success and been extensively used in $r-$process calculations over the years. However, despite the success of the mic-mac models in fitting experimental data on the measured masses of nuclei, the root mean square (rms) errors with respect to the known experimental data are not at the desired level. The rms error, in relation to the available mass data, was found to be 0.298 MeV when using the WS4 model \cite{WANG2014215} and 0.662 MeV using the FRDM(2012) model \cite{MOLLER20161}. Furthermore, these models have been fitted using the available experimental data, which makes the behavior of nuclei on the neutron-rich side of the nuclear chart still somewhat uncertain.

More sophisticated methods, such as self-consistent mean-field (SCMF) theories based on the Hartree-Fock-Bogolyubov (HFB) approach with nuclear energy density functionals (EDF), have also long been employed to investigate the properties of nuclei. Although using microscopic tools in calculations is a computationally demanding task, large-scale computations of the nuclear chart are nonetheless available. In recent years, both relativistic and non-relativistic calculations have been performed using different EDFs to probe the properties of nuclei and to define the boundaries of the nuclear landscape \cite{ZHANG2022101488,XIA20181,PhysRevC.93.054310,PhysRevC.91.014324,PhysRevC.89.054320,Erler2012,Grams2023, PhysRevC.108.054305,ravlic2023global,Ravlic2023}. While these models perform well around the stability line with respect to the available experimental data, they reveal local variations and significant discrepancies that increase with neutron number, ultimately impacting the location of the drip lines. The major source of these differences is the missing incomplete correlations on the purely mean-field level of the HFB description, the usage of different interactions that are optimized using different strategies, pairing correlations, and the impact of the continuum. Typically, the rms error of these mass tables compared to the available experimental data are quite high and range between 2.0 and 5.0 MeV, depending on the interaction used in the calculations. One of the most recent functionals, which are optimized using the experimental data of all available nuclei, has reached an rms error between 0.5 and 0.6 MeV \cite{PhysRevC.88.024308, PhysRevC.96.044308, PhysRevC.88.061302, PhysRevC.93.034337, Grams2023}. Considering all of these factors, there is a need to find fast and reliable methods for determining nuclear properties, especially away from the stability line.

In recent years, machine learning (ML) models have gained considerable attention within the scientific community, demonstrating notable success, including also the field of nuclear physics (see Ref. \cite{RevModPhys.94.031003} and references therein). These models have proven capable of directly predicting nuclear properties using experimental data \cite{GAZULA19921,CLARK_2006, ATHANASSOPOULOS2004222, yuksel2021, BAHTIYAR2022109470, PhysRevC.105.064306, universe7050131, PhysRevC.106.014305, PhysRevC.106.L021301,PhysRevC.105.064306,10.3389/fphy.2023.1198572}. Recently, the ML models have also been used to improve mic-mac and microscopic model predictions. Within this framework, the most popular tool is Bayesian neural networks (BNNs), which have been used to improve the results of microscopic calculations by training on the residuals. These residuals represent the differences between experimental data and microscopic calculations, and the BNNs have gained considerable attention and success in that respect \cite{PhysRevC.97.014306,NIU201848, PhysRevC.98.034318, PhysRevLett.122.062502}. In this context, ML models can be used as reliable and efficient tools to probe nuclear properties; however, more studies are necessary to better understand their predictive capability.

In this study, our goal is to assess the performance of the two ML models in predicting the nuclear mass excess ($M$) of nuclei, rather than correcting existing mic-mac or microscopic model predictions. We use the Support Vector Regression (SVR) and Gaussian Process Regression (GPR) ML models to calculate the mass excess of nuclei. These models are trained using available experimental data along with the relevant physics-based feature space. Then, we evaluate the performance of these ML models in predicting the mass excess of nuclei, examining also their extrapolation capabilities far beyond the training and test regions.

\section{Machine Learning models}
In this section, we present an overview of the ML models employed in our calculations: SVR and GPR. Additionally, we describe the experimental data used to train these models and provide details about the physics-based feature space involved.

\subsection{Support Vector Regression}

The SVR \cite{drucker1996} is an ML model specifically designed for tackling regression tasks, offering a unique approach to predict continuous outcomes by leveraging the principles of support vector machines (SVMs) \cite{boser1992}. In contrast to classification-focused methods, SVR seeks a hyperplane that optimally fits the training data with minimized error. Central to SVR is the concept of \emph{support vectors}, which are critical data points closest to the hyperplane’s boundaries. The model aims to align as many data points as possible within the optimal hyperplane, fitting within a specified tolerance margin. It simultaneously controls margin violations, addressing instances where data points exceed the boundaries. This brings a hyperparameter $\epsilon$, which controls the width of the hyperplane \cite{geron2017}. 

The strength of SVR is particularly notable in addressing non-linear regression problems, often yielding enhanced results \cite{scholkopf2002}. At the core of SVR's approach to these problems is the effective \emph{kernel trick}. This technique is crucial when dealing with input data that is not linearly separable in its original feature space. By employing the kernel trick, SVR can implicitly project the data into a higher-dimensional space, achieving linear separability. This projection is facilitated by a kernel function, which efficiently calculates the dot product of data point pairs in this higher-dimensional space without the need for explicit calculation of transformed features. Thus, SVR involves mapping input data into a higher-dimensional space using kernel functions, allowing for the capture of nonlinear relationships. Among various kernel functions, the Radial Basis Function (RBF) kernel is a widely used choice in SVR applications defined as:

\begin{equation}
\begin{aligned}
K_G(x, x') &= \exp\left(-\frac{||x - x'||^2}{2\sigma^2}\right) \\
         &= \exp\left(-\gamma ||x - x'||^2\right).
\end{aligned}
\label{kernel}
\end{equation}

In this context, $x$ and $x'$ represent two data points, and their Euclidean distance is denoted as $||x - x'||$, while $\gamma$ is the kernel coefficient. Eq. \ref{kernel} quantifies the similarity or dissimilarity between these data points, based on their distance in the input feature space. This results in higher similarity for closer data points, and conversely, lower similarity for those more distant \cite{geron2017, soydaner2023}. The tuning of hyperparameters plays a crucial role in SVR. Proper parameter adjustment is important to prevent overfitting or underfitting, ensuring the model generalizes well to unseen data. The regularization hyperparameter, denoted as $C$, is essential in striking the balance between maximizing the margin and minimizing the training error. Additionally, the $\epsilon$ hyperparameter is important in determining the tolerance margin, within which the epsilon-insensitive loss function does not penalize errors. Data points within this margin do not contribute to the loss function, enhancing the model's robustness against minor prediction errors and improving its resilience to outliers. In our experiments, we set the $\epsilon$ to 0.002, $C$ to 1000, and $\gamma$ to 0.03. Another hyperparameter, the \emph{tolerance}, which indicates the desired precision for convergence, is set to $10^{-5}$. We performed calculations on several hyperparameter configurations to determine the optimal setting for our task, and ultimately report the model that exhibits the highest performance.

SVR offers a versatile framework for regression tasks, utilizing kernels to capture diverse relationships and incorporating a margin of tolerance to enhance robustness. Practical hyperparameter tuning and understanding the role of kernels are fundamental for optimizing the model performance across various datasets and maximizing its efficiency in real-world applications.

\subsection{Gaussian Process Regression}

When we consider a linear model expressed as $y = w^T x$, this model describes a linear relationship for every different value of $w$. If we introduce a prior distribution for $w$, denoted as $p(w)$, the distribution of possible $y$ values at any given $x$, $y(x|w)$, emerges from sampling $w$ from $p(w)$. This is the main idea of a \emph{Gaussian process}. When $p(w)$ follows a Gaussian distribution, each resulting $y$ is also Gaussian, being a linear combination of Gaussians. Specifically, our interest lies in the joint Gaussian distribution of $y$ values computed at $N$ input data points $x^t$, where $t = 1,\dots,N$ \cite{mackay1998,alpaydin2014}. We typically assume a Gaussian prior with zero-mean for $w$, as shown in Eq.\ref{gaussianprocess}: 

\begin{equation}
p(w) \sim \mathcal{N}(0, (1/ \alpha) I).
\label{gaussianprocess}
\end{equation}

GPR \cite{mackay1998, rasmussen2006} operates by leveraging Gaussian processes to model distributions over functions. Initially, the algorithm establishes a prior distribution over functions, assuming Gaussian-distributed function values at input points. This prior distribution forms the foundation, characterized by a mean function and a kernel function. As training data is observed, this prior is updated to a posterior distribution using Bayes' theorem. This update incorporates the observed data, refining the model's beliefs about the underlying function. The resulting posterior distribution enables predictions at new data points, providing not only a mean prediction but also an associated measure of uncertainty, which is crucial for decision-making in uncertain scenarios.
 
The choice of kernels is important in GPR. In our work, we utilize a combination of the RBF kernel (Eq.\ref{kernel}) and the White kernel (Eq.\ref{whitekernel}). The RBF kernel is particularly effective at capturing intricate data patterns, adapting to various scales, and ensuring smooth connections between data points. Meanwhile, the White kernel models noise within the dataset. Adjusting the noise level hyperparameter within the White Kernel is essential, striking a balance between capturing the underlying signal and accommodating inherent noise. This delicate interplay between kernels enables our GPR model to provide robust predictions, while acknowledging and quantifying uncertainties. To optimize the model, we adjust the kernel parameters by spanning a range of values to obtain the optimum values for our calculations. In our experiments, the length scale of the RBF kernel is set to 1.0, with its lower and upper bounds on the length scale being ($10^{-4}$, $10^{5}$). For the White Kernel, the noise level is set at 1, with the noise level's lower and upper bounds set at ($10^{-10}$, $10$).

\begin{equation}
K_W(x, x') = \begin{cases} 
\text{noise level} & \text{if } x_i = x'_j \\
0 & \text{otherwise}.
\end{cases}
\label{whitekernel}
\end{equation}

GPR's strength lies not only in its predictive accuracy but also in its ability to provide nuanced insights into the reliability of those predictions. This is achieved through its uncertainty quantification, which offers a probabilistic measure of confidence in its predictions. By effectively quantifying the uncertainty associated with each prediction, GPR enhances decision-making processes in various domains. This dual capability of delivering precise predictions while simultaneously assessing their reliability makes GPR a valuable tool across a wide range of applications.

\begin{figure} [ht!]
 \centering
\includegraphics[width=\linewidth]{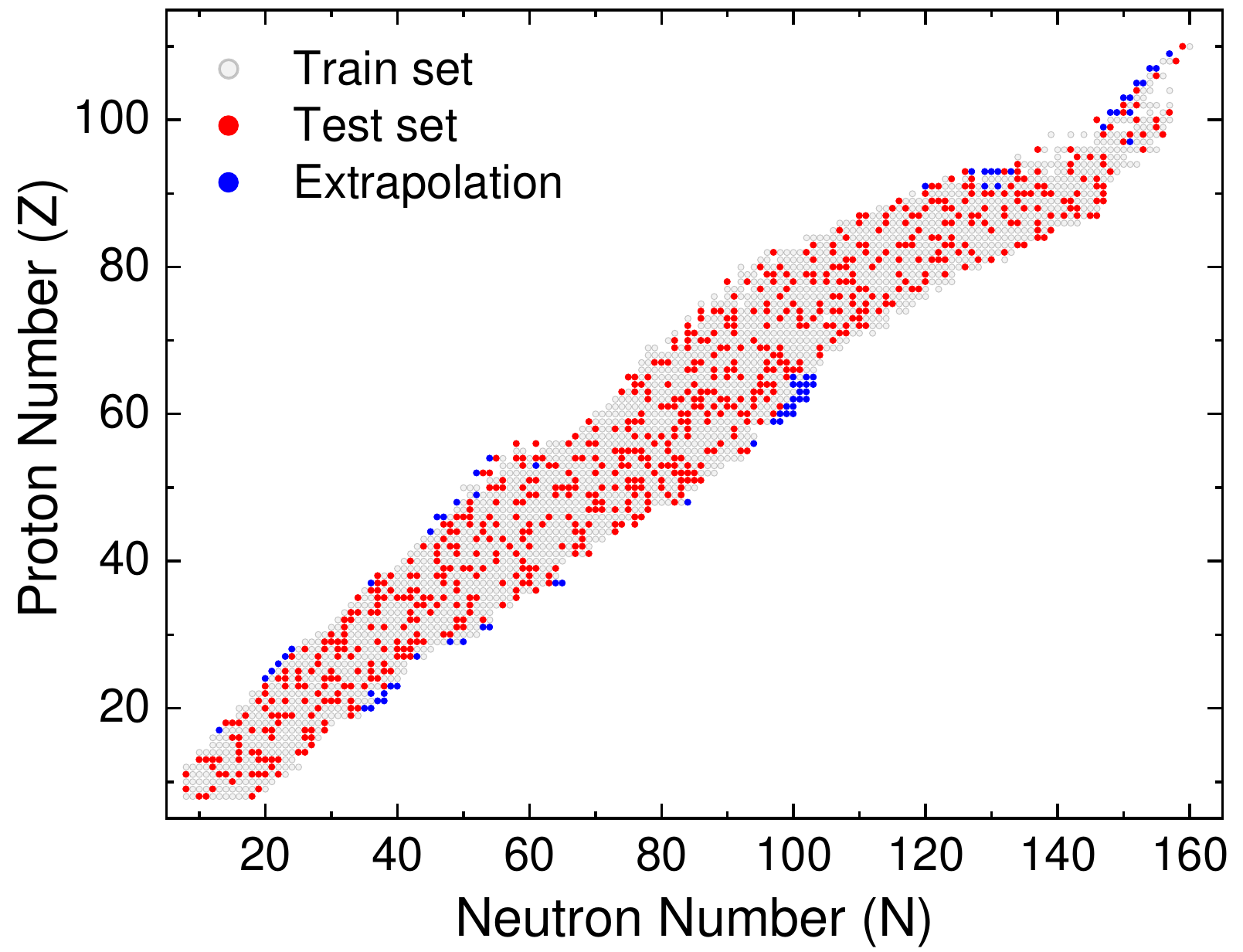} 
\caption{The training set (gray circles), test set (red circles), and extrapolation set (blue circles) used in the ML models. Both the training and test sets include nuclei from AME2020, with the exception of the newly measured 71 nuclei from AME2020, which are exclusively designated for the extrapolation set \cite{Wang_2021}.}\label{fig:2}
\end{figure}

\subsection{The experimental data and feature space}
In this study, our objective is to develop an ML model that predicts the mass excess of atomic nuclei using both experimental data on the mass excess of nuclei and a physics-based feature space. The experimental mass excess values are taken from the atomic mass evaluation 2020 (AME2020) \cite{Wang_2021} for nuclei with $Z, N \geq 8$ (2386 nuclei). Then, the experimental data is randomly divided into two subsets: $75.0\%$ (1789 nuclei) for training and $25.0\%$ (597 nuclei) for testing. The nuclei in the training and test sets remain the same for all calculations. The performance of the ML models is also assessed beyond the training and test data sets. The AME2020 data includes new experimental information for 71 nuclei compared to the previous AME2016 \cite{Wang_2017}. These nuclei have been utilized to test the extrapolation capabilities of the models. The estimated mass excess values, derived from the trends in the mass surface (TMS) of nuclei, are also utilized to compare our findings in the extrapolation region. The selection of the training and test sets, as well as the new data from AME2020 (extrapolation set), is shown in Figure \ref{fig:2}. Additionally, we evaluate the extrapolation performance of the models by extending calculations to the neutron-rich region beyond the reach of current experimental facilities, probing the limits of their predictive capabilities.

\begin{figure*} [ht!]
 \centering
\includegraphics[width=\linewidth]{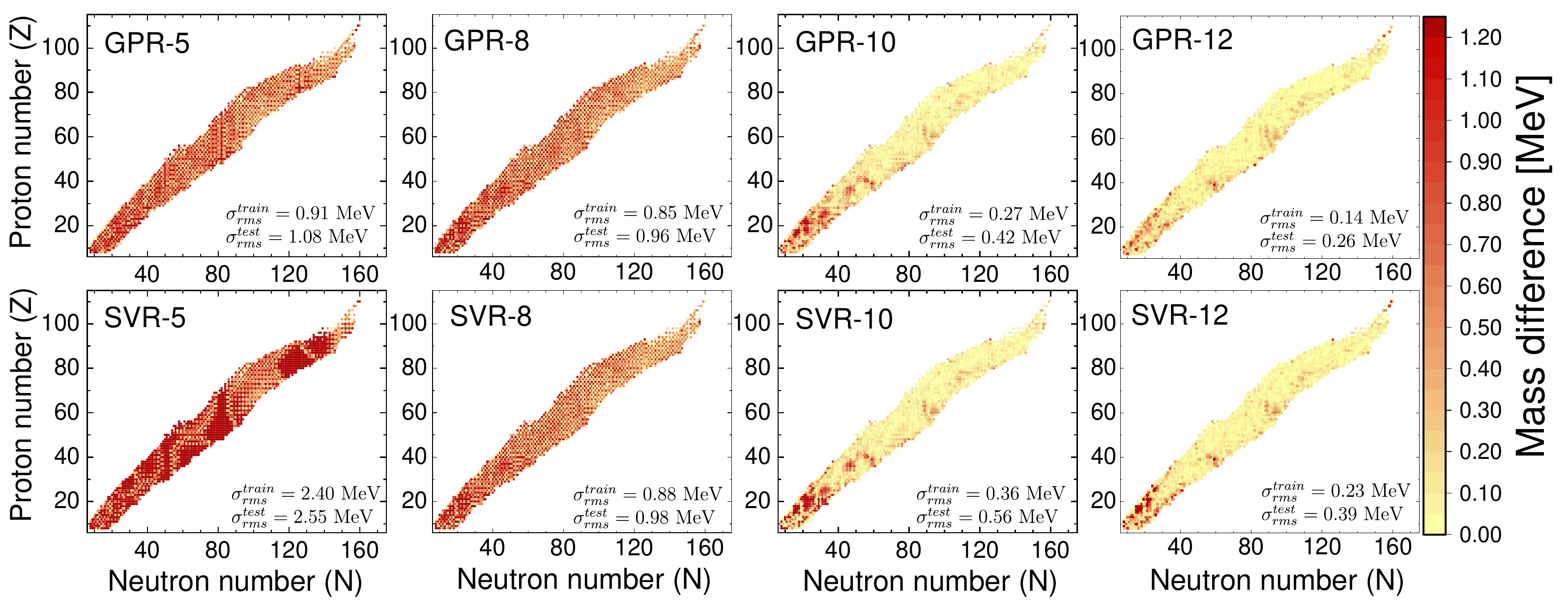} 
\caption{The absolute value of mass excess differences between the GPR and SVR predictions for training and test set using different features (see Table \ref{tab:trainingSets}) and the AME2020 data \cite{Wang_2021}. The rms errors for the training and test sets are also provided.}\label{fig:3}
\end{figure*}

\begin{table}[ht!]
\centering
\renewcommand\arraystretch{1.5}
\setlength{\tabcolsep}{10pt} %
\caption{The ML models with different features.}
\begin{tabular}{ccc}
\hline \hline \textbf{Model} & \textbf{Feature Space}  \\ \hline
SVR/GPR-5 & $Z$, $N$, $A$, $A^{2/3}$, $(N-Z)/A$  \\ 
SVR/GPR-8 & $Z$, $N$, $A$, $A^{2/3}$, $(N-Z)/A$, \\
          &     $\nu_{Z}$, $\nu_{N}$, $PF$ \\
SVR/GPR-10& $Z$, $N$, $A$, $A^{2/3}$, $(N-Z)/A$, \\
          & $\nu_{Z}$, $\nu_{N}$, $PF$, $Z_{eo}$, $N_{eo}$\\
SVR/GPR-12& $Z$, $N$, $A$, $A^{2/3}$, $(N-Z)/A$, \\
          & $\nu_{Z}$, $\nu_{N}$, $PF$, $Z_{eo}$, $N_{eo}$, $Z_{shell}$, $N_{shell}$\\
\hline \hline 
\end{tabular}
\label{tab:trainingSets}
\end{table}

As it is well known, the use of appropriate inputs during training can significantly impact the performance of ML models \cite{PhysRevC.105.064306, PhysRevC.106.014305, PhysRevC.106.L021301, yuksel2021, BAHTIYAR2022109470}. Therefore, in our models, we incorporate relevant features of nuclei that can influence mass predictions. Our feature space consists of 12 inputs: $Z$, $N$, $A$, $A^{2/3}$, $(N-Z)/A$, $Z_{eo}$, $N_{eo}$, $\nu_{Z}$, $\nu_{N}$, $PF$, $Z_{shell}$, and $N_{shell}$.
Here, the bulk properties are defined as the proton (neutron) number $Z$ ($N$), the mass number ($A$), and $A^{2/3}$ for volume and surface terms. The term $\frac{N - Z}{A}$ is a measure of isospin asymmetry. The odd-even nature of protons ($Z_{eo}$) and neutrons ($N_{eo}$) is defined as follows: $Z_{eo}$ ($N_{eo}$) equals zero when $Z$ ($N$) is even and one when $Z$ ($N$) is odd. We also provide information about the nuclear magic numbers: $\nu_{Z}$ and $\nu_{N}$ represent the valence number of protons and neutrons measured from the nearest closed shell. The nuclear magic numbers for protons and neutrons are taken as $Z(N) = 8, 20, 28, 50, 82, 126, 184$. The promiscuity factor ($PF$) is represented by the formula $PF = \frac{\nu_{Z} \cdot \nu_{N}}{\nu_{Z} + \nu_{N}}$, and serves as a measure of valence proton-neutron $(p - n)$ interactions \cite{Casten_1996}. Lastly, the system is informed about the nuclear shells with $Z_{\text{shell}}$ and $N_{\text{shell}}$; they represent the shell model orbitals of the last proton and neutron. The values of $Z_{\text{shell}}$ and $N_{\text{shell}}$ are defined as 0, 1, 2, 3, or 4, depending on whether the proton or neutron number falls within the specified ranges: 1–28, 29–50, 51–82, 83–126, and above 127, respectively \cite{VOGT2001255}. In order to assess the importance of the feature space in model calculations, we implement ML models with different features. The inputs used in our ML models are given in Table \ref{tab:trainingSets}.

\section{Results}

Figure \ref{fig:3} displays the absolute differences between the results of GPR (upper panels) and SVR (lower panels) with different inputs and the experimental data taken from AME2020 \cite{Wang_2021}. The feature space of the ML models is provided in Table \ref{tab:trainingSets}. The rms errors for the training and test sets of each selected model are also presented in Figure \ref{fig:3}. Using only the bulk properties of nuclei to construct the model, GPR-5 yields reasonable results, with rms errors of 0.91 and 1.08 MeV for the training and test sets, respectively, better than most of the microscopic model calculations. On the other hand, the performance of SVR-5 is lower compared to GPR-5, with rms errors of 2.40 and 2.55 MeV for the training and test sets, respectively.

In Figure \ref{fig:3}, it is evident that increasing the physics-based feature space significantly improves the performance of the models. The importance of the physics-based feature space has also been discussed in previous studies, with similar results obtained using different ML models \cite{yuksel2021,PhysRevC.106.L021301,PhysRevC.106.014305,PhysRevC.105.064306}. It has been noticed that the inclusion of the odd-even nature of protons and neutrons ($Z_{eo}$, $N_{eo}$) leads to a significant improvement in ML predictions. Subsequently, the results improve further with the inclusion of information on the nuclear shells, $Z_{\text{shell}}$ and $N_{\text{shell}}$. Utilizing 12 inputs (GPR-12) in the calculations, we achieved an rms error of 0.14 and 0.26 MeV for the training and test sets, respectively. These rms error values are even better than those of well-known mic-mac mass models, suggesting that the GPR model effectively captures the given information of nuclei and makes reasonable predictions. Additionally, we observe that GPR performs better for medium-heavy and heavy nuclei, while errors are slightly higher for light nuclei. The poorer performance in light nuclei is attributed to the lower number of available experimental data in this region. Similar results are also obtained using the SVR model. However, we find that SVR requires more data and information to make reasonable predictions for training and test set nuclei, and GPR outperforms SVR in that respect.

We also compare our findings with previous ML studies in which different ML models have been used to predict nuclear mass excess. One of the first applications of ML models in nuclear physics was performed using SVMs \cite{CLARK_2006}, predicting nuclear mass excess long ago. It yielded rms errors of 0.35, 0.5, and 0.71 MeV for the training, validation, and test sets, respectively. A recent application of the probabilistic ML algorithm, the Mixture Density Network (MDN), has yielded rms errors of around 0.5-0.6 MeV with respect to the AME2016 \cite{PhysRevC.106.014305} when supplemented with physics-based feature space. Recently, it has been shown that the inclusion of a soft physical constraint in the MDN achieved an rms error of 0.186 MeV for the training data (consisting of only 450 nuclei, approximately $20\%$ of the AME2016 dataset) and an rms error of 0.316 MeV for the remainder of the AME2016 data with $Z\geq 20$ \cite{PhysRevC.106.L021301}. Therefore, we also performed calculations using different train-test set ratios to assess the performance of our ML models, and the results are presented for nuclei with $Z\geq 8$ in Table \ref{ratio}. Our findings indicate that our ML models exhibit robust predictive capabilities even when trained on a mere $25\%$ of the available experimental data. However, as expected, the models' performance on the test set declines with reduced training data, as they struggle to grasp details with limited information. Conversely, increasing the number of the training data yields noticeable improvements in the models' test set performance, while the performance on the training set remains relatively stable. Similar results are also obtained in Ref. \cite{10.3389/fphy.2023.1198572} using the MDN, whereas it is observed that our ML models require more training data to learn and generalize information to unseen data compared to the MDN. We anticipate that incorporating physical constraints into ML models, such as the Garvey-Kelson (GK) relations, can also enhance the predictive power of the ML model on unseen data, particularly with a limited amount of training data \cite{PhysRevC.106.L021301}. Alternatively, increasing the size of the training data, as demonstrated in our work, can also improve model performance on unseen data. Our results, even without applying a physical constraint, are in good agreement with the findings in Refs. \cite{PhysRevC.106.L021301,10.3389/fphy.2023.1198572} when using train-test set ratios of $50\%$-$50\%$ and $75\%$-$25\%$.

\begin{table}[ht!]
    \centering
    \renewcommand\arraystretch{1}
    \setlength{\tabcolsep}{5pt} %
    \caption{Root mean square errors $\sigma_{rms}$ (in MeV) for GPR-12 and SVR-12 ML models, indicating their performance on training and test sets for $Z\geq 8 $ across varying train-test data ratios from AME2020 set \cite{Wang_2021}. The percentages represent the proportion of data allocated to the training and test sets.}
    \begin{tabular}[t]{lccc}
        \hline \hline
       && \multicolumn{1}{c}{Train-test ratio $\%$} \\
       \hline
        & \multicolumn{1}{c}{25-75} & \multicolumn{1}{c}{50-50} & \multicolumn{1}{c}{75-25} \\
        \hline
         GPR-12 (train) & 0.16 & 0.21 & 0.14  \\[1mm]
         GPR-12 (test) & 0.79 & 0.49 & 0.26  \\[1mm]
        \hline
         SVR-12 (train) & 0.13 & 0.20 & 0.23 \\[1mm]
        SVR-12 (test) & 0.91 & 0.49 & 0.39 \\[1mm]
        \hline 
        \label{ratio}
    \end{tabular}
\end{table}

We conclude that our findings, obtained using different ML models, not only align with these previous studies but also establish GPR and SVR as alternative and reliable tools for ML studies in nuclear physics.

\textit{Extrapolation performance of ML models - } One of the most important issues in ML studies is the low performance of the ML models when it comes to extrapolation, namely, outside the training and test set regions. It is essential to develop models that not only predict well-known experimental data (training and test data) effectively but can also make accurate predictions for parts of the nuclear chart that are challenging to measure experimentally. Therefore, in this subsection, we assess the extrapolation capabilities of the ML models by extending beyond the experimentally known region. Initially, we test the performance of the ML models on the newly measured 71 nuclei from the AME2020 data \cite{Wang_2021} (see Fig. \ref{fig:2}). We present the rms errors of each model in Table \ref{extra}. Clearly, the accuracy of model predictions improves with the use of appropriate features. Specifically, increasing the number of inputs from 5 to 12 improves the performance of the GPR and SVR models in the extrapolation region by $54.73\%$ and $67.96\%$, respectively. Furthermore, the low rms errors of these ML models, which are comparable to those of modern nuclear mass models, indicate that ML models are able to make reasonable predictions even outside the training region.

\begin{table}[th!]
    \centering
    \renewcommand\arraystretch{1}
    \setlength{\tabcolsep}{5pt} %
    \caption{The root mean square errors (given in MeV) for the extrapolation set (71 nuclei from AME2020). The calculations are performed using different inputs.}
    \begin{tabular}[t]{lcccc}
        \hline \hline
        \multicolumn{1}{c}{Feature} & \multicolumn{1}{c}{5} & \multicolumn{1}{c}{8} & \multicolumn{1}{c}{10} & \multicolumn{1}{c}{12} \\
        \cmidrule(lr){1-1}
        \cmidrule(lr){2-2}
        \cmidrule(lr){3-3}
        \cmidrule(lr){4-4}
        \cmidrule(lr){5-5}
        \multicolumn{1}{c}{Model}   & $\sigma_{rms}^{extrap.}$ &$\sigma_{rms}^{extrap.}$ & $\sigma_{rms}^{extrap.}$& $\sigma_{rms}^{extrap.}$ \\
        \hline
        GPR & 1.48 & 1.10 & 0.75 & 0.67  \\[1mm]
        SVR & 2.31 & 1.17 & 0.70 & 0.74 \\[1mm]
        \hline \hline
        \label{extra}
    \end{tabular}
\end{table}

\begin{figure*} 
 \centering
\includegraphics[width=\linewidth]{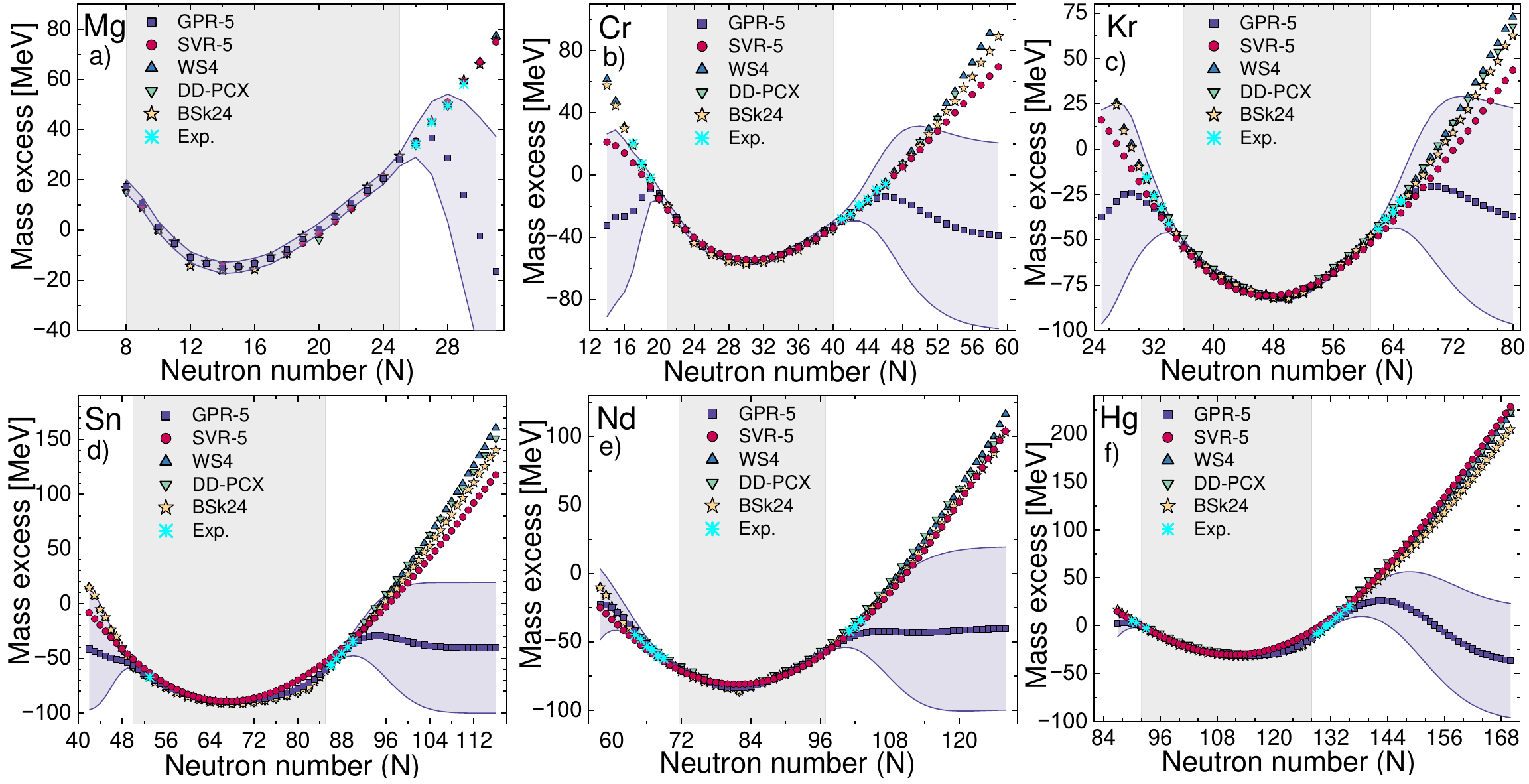} 
\caption{The GPR-5 and SVR-5 mass excess predictions are shown for the selected isotopic chains as a function of the neutron number. The blue shaded region represents the $95.0\%$ confidence interval, and the gray region indicates the training and test set area, while the white region is used as extrapolation region. The estimated values for the mass excess predictions, away from the training and test set region, are derived from the trends in the mass surface (TMS) and are taken from Ref. \cite{Wang_2021}. Predictions of other mass models: mic-mac model WS4 \cite{WANG2014215}, non-relativistic Skyrme-type BSk24 interaction \cite{PhysRevC.88.061302}, and relativistic point-coupling interaction DD-PCX \cite{PhysRevC.99.034318, PhysRevC.108.054305}, are also provided for comparison.}\label{fig:4}
\end{figure*}

\textit{How far can we go from the experimentally known region and get reasonable results using ML models?} In order to assess the extrapolation performance of the ML models, we extend our calculations through the proton-rich and neutron-rich regions. The results are presented for both the training and test regions (gray region) and the extrapolation region (white regions), where no experimental data currently exists. In Figure \ref{fig:4}, we depict the predictions for the mass excess of nuclei using GPR-5 and SVR-5 for selected isotopic chains from various parts of the nuclear chart. The estimated values for the mass excess predictions from the trends in the mass surface (TMS) are also used to assess the performance of the models in the extrapolation region  \cite{Wang_2021}. Additionally, we compare these predictions with results from well-known mass tables: the mic-mac model WS4+RBF \cite{WANG2014215} and the non-relativistic (BSk24) \cite{PhysRevC.88.061302} calculations. The relativistic calculations with the point-coupling interaction DD-PCX \cite{PhysRevC.99.034318} are performed for even-even nuclei using the axially-deformed Hartree-Bogoliubov (RHB) model with separable pairing \cite{NIKSIC20141808}, employing 20 harmonic oscillator shells for convergence in the calculations \cite{PhysRevC.108.054305}.

\begin{figure*} 
 \centering
\includegraphics[width=\linewidth]{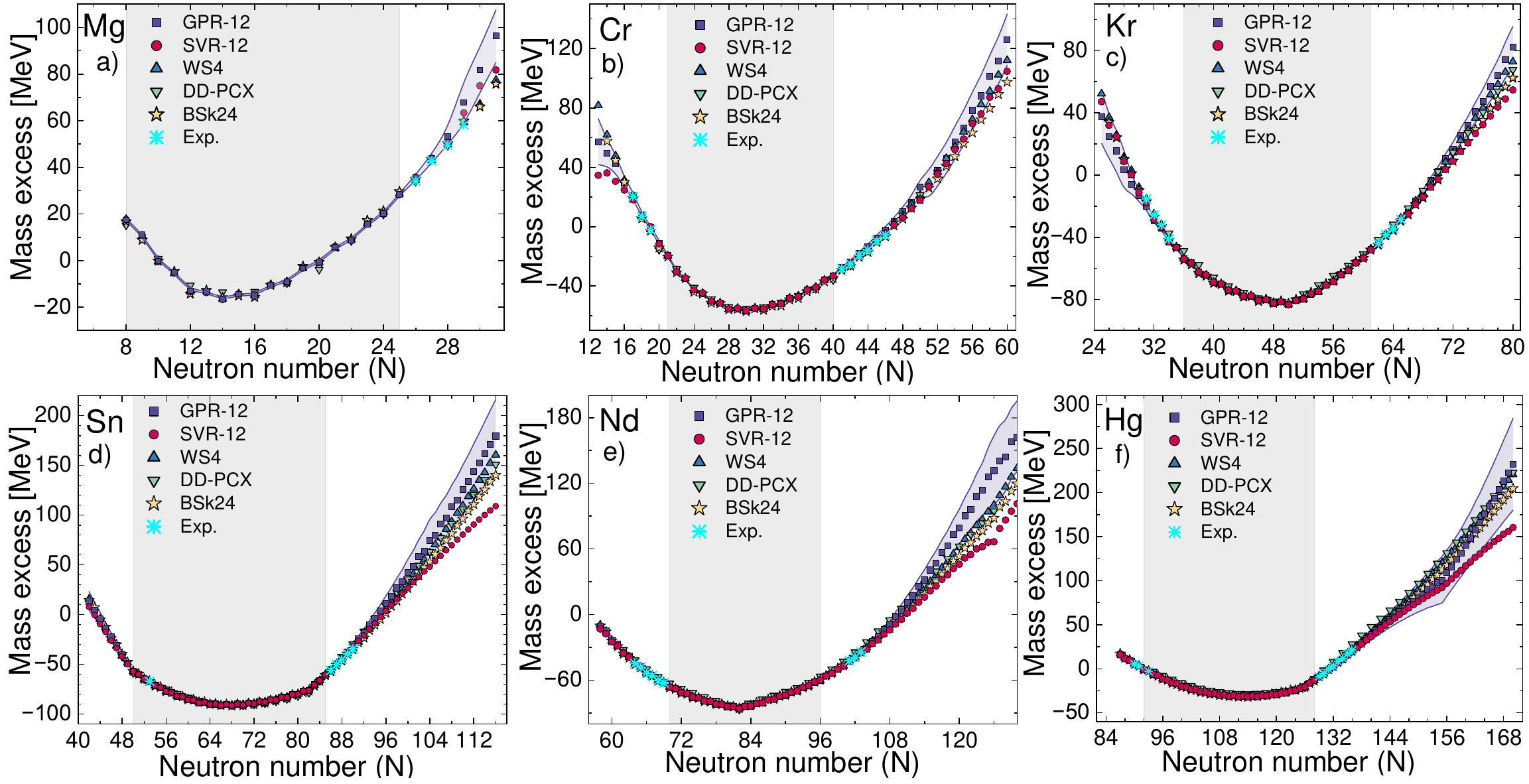} 
\caption{The same as in Fig. \ref{fig:4} but using GPR-12 and SVR-12 ML models.}\label{fig:5}
\end{figure*}

\begin{figure*} 
 \centering
\includegraphics[width=0.8\linewidth]{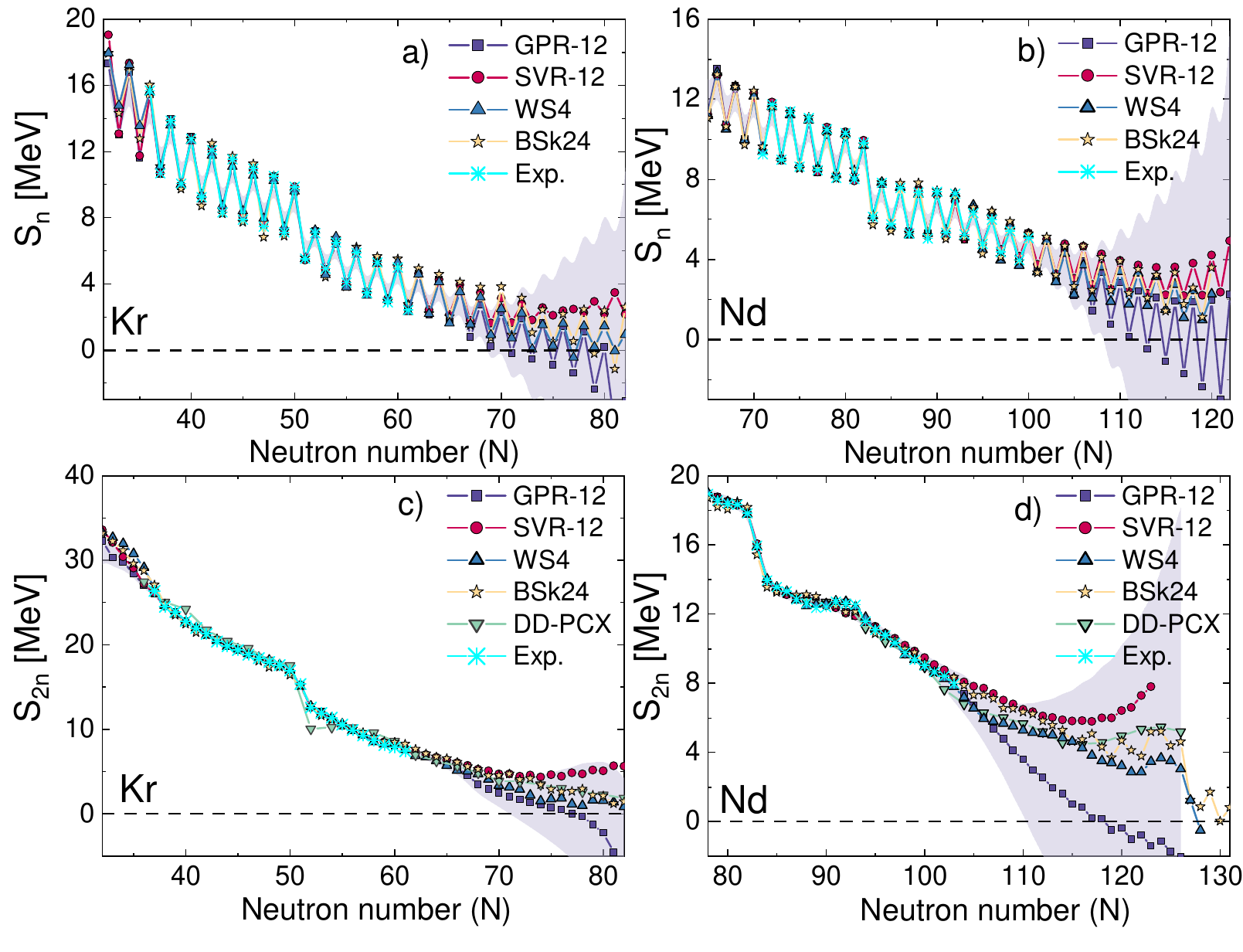} 
\caption{Upper panels: one-neutron separation energies for Kr (a) and Nd (b) isotopic chains using GPR-12 and SVR-12 models. Lower panels: two-neutron separation energies for Kr (c) and Nd (d) isotopic chains. The blue shaded region represents the $95.0\%$ confidence interval. Theoretical model calculations (WS4, BSk24, DD-PCX) and experimental data are also provided when available \cite{Wang_2021}.}\label{fig:6}
\end{figure*}

In GPR, the uncertainty is represented by the blue shaded region. It represents the probability distribution over the possible functions. This distribution is updated as more data or features are observed, which leads to a more precise estimate of the function. Therefore, it is expected that the uncertainty increases away from the training data, which is a direct consequence of the roots of Gaussian Process in probability and Bayesian inference. As can be seen from the upper panels of Figure \ref{fig:4}, the GPR with only 5 features performs poorly when we move away from the training-test region, and the uncertainty is quite high in the extrapolation region. Apart from the Mg chain, the GPR can make reasonable predictions for the isotopic chains up to an increase in neutron number around 4 or 5. Then, the results start to deviate and do not follow the trends obtained in different mass models. Although the rms errors are higher for the training and test sets using the SVR-5 model, it is seen that the SVR-5 model captures the trends better in the extrapolation region.

By increasing the number of features in the GPR model, we observe a significant improvement in the model's performance in the extrapolation region (Figure \ref{fig:5}). Firstly, we note a considerable reduction in the uncertainties of the predictions. Secondly, the predictions of the GPR-12 model align with a trend that is similar and comparable to those obtained in different mass models, albeit slightly higher nearby the drip line. Increasing the number of features in the GPR model unequivocally enhances its generalizability and improves uncertainty estimation. As mentioned above, SVR-12 demonstrates improved predictions in both the training and test regions when the number of features is increased. However, an increase in the number of features in the SVR model does not lead to better results for the extrapolation region. The predictions of the SVR models start to deviate from other mass models and underestimate the mass excess values compared to them near the drip lines.

Finally, we explore the one- and two-neutron separation energies calculated using the mass excess $M$ values obtained from our ML models and compare them with those from other models and available experimental data. The one and two neutron separation energies are calculated by

\begin{equation}
\begin{split}
 S_{n} &= - M(A,Z) + M(A-1,Z) + m_{n}, \\
 S_{2n} &= - M(A,Z) + M(A-2,Z) + m_{2n},
\end{split}
\end{equation}
where $m_{n}$ represents the mass of the neutron.
In the upper panels of Figure \ref{fig:6}, the results are displayed for the one-neutron separation energies of Kr (a) and Nd (b) isotopic chains. It is evident that the ML models provide reasonable predictions and are in agreement with the experimental data, exhibiting the well-known odd-even staggering (OES) in binding energies. As the neutron number increases, the results also show comparability with other theoretical model calculations. However, near the drip lines, the ML models start to deviate from other model calculations.

In the lower panels of Fig. \ref{fig:6}, the two-neutron separation energies are displayed for the Kr (c) and Nd (d) isotopic chains. It can be observed that the ML models make reasonable predictions for the Kr chain. In comparison to the SVR-12 model, the GPR-12 model's predictions are more reasonable near the drip lines and follow a smooth decreasing behaviour with increasing neutron number. Additionally, the predictions of the GPR-12 model are comparable to the WS4 model, while the SVR-12 model results align with the BSK4 model as neutron number increases. When it comes to nuclei near the drip lines, the predictions of the SVR-12 model become inaccurate and exhibit an increasing pattern. The ML model predictions deviate from other mass models, particularly for heavier Nd nuclei. It is also seen that the uncertainty in the GPR-12 predictions is higher for this chain in the extrapolation region. This discrepancy is a natural consequence of both the limited number of available experimental data points and the absence of information in the physics-based feature space in this particular region.

\textit{Do the results of the ML models satisfy the Garvey-Kelson mass relations?} The Garvey-Kelson relations \cite{RevModPhys.41.S1}, which are based on the independent particle shell model, consist of mathematical expressions that establish links among the masses of neighboring nuclides. These relations arise from the condition that various interactions between nucleons cancel out at the first order, resulting in a series of mass relations between adjacent nuclei \cite{RevModPhys.41.S1, PhysRevC.77.041304}. The GK mass relation for nuclei with $N\geq Z$ is given by 

\begin{equation}
\begin{split}
&M(Z-2,N+2) - M(Z,N) \\
&+ M(Z-1,N) - M(Z-2,N+1) \\
&+ M(Z,N+1) - M(Z-1,N+2) \approx 0,
\end{split}
\label{gk1}
\end{equation}
and for nuclei with $Z<N$, 

\begin{equation}
\begin{split}
&M(Z+2,N-2) - M(Z,N) \\
&+ M(Z,N-1) - M(Z+1,N-2) \\
&+M(Z+1,N) - M(Z+2,N-1) \approx 0.
\end{split}
\label{gk2}
\end{equation}

\begin{figure} 
 \centering
\includegraphics[width=\linewidth]{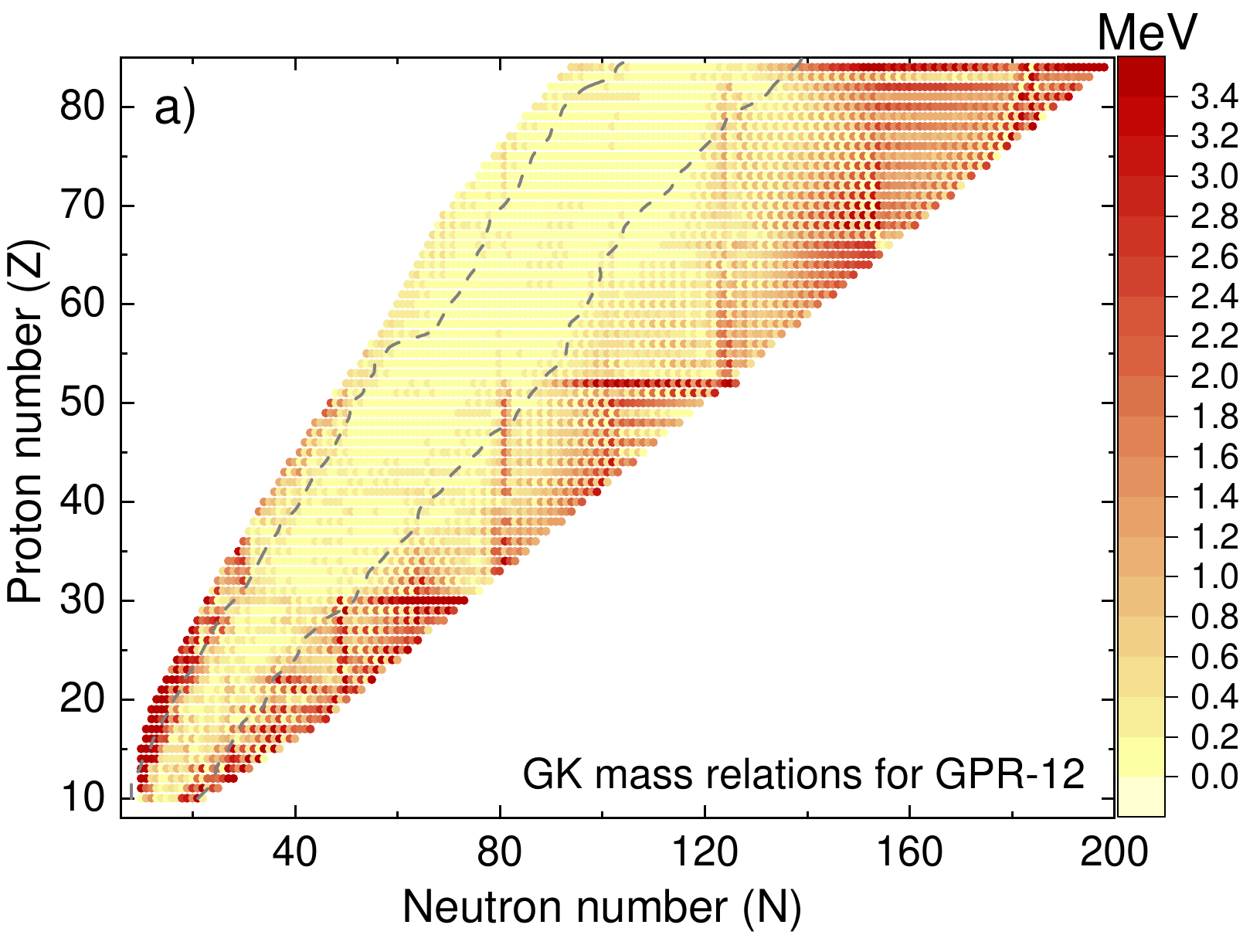} 
\includegraphics[width=\linewidth]{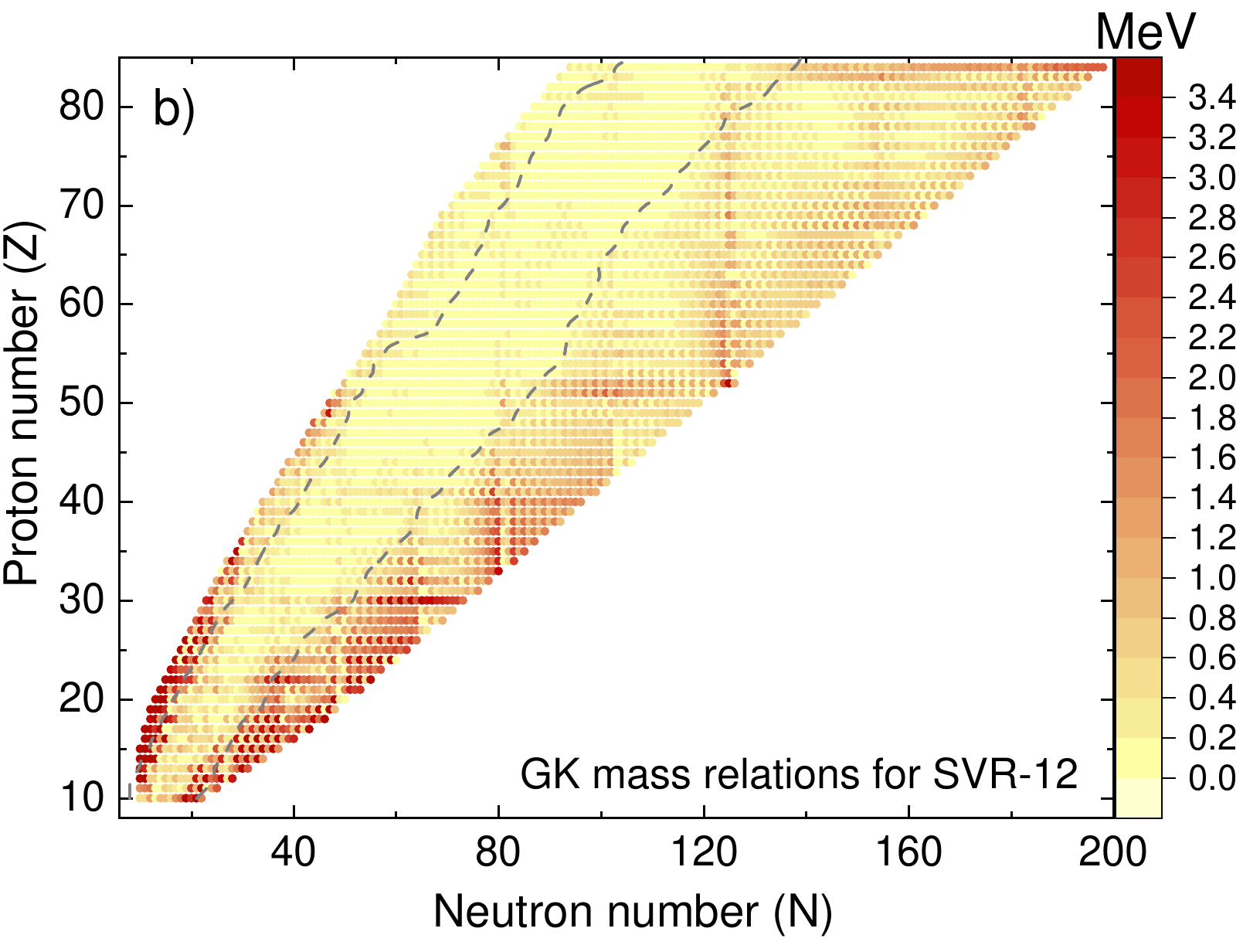} 
\caption{The GK mass relations for the results obtained using the (a) GPR-12 and (b) SVR-12 ML models. The dashed gray lines indicate the borders of the training and test set regions.}\label{fig:66}
\end{figure}

Using the results obtained from the GPR-12 and SVR-12 models, we also assess whether the GK relations are maintained in our ML models, serving as an additional evaluation of the ML models and their extrapolation abilities. In Fig. \ref{fig:66}, we present the results of the GK relationship described by Eqs. \ref{gk1} and \ref{gk2}. It is evident that the GK relationships are well maintained within the training and test set regions for the ML models under consideration. However, deviations become apparent with increasing proton and neutron numbers, especially for low mass nuclei and throughout the neutron drip lines. Interestingly, while the GPR-12 model seems to perform better than the SVR-12 model near the neutron drip line (see Fig. \ref{fig:6}), we find that the SVR-12 model exhibits better performance in the neutron-rich region concerning the GK mass relations. 
The differences between GPR and SVR predictions can be attributed to their distinct mathematical principles and model complexities. Including physical constraints, such as GK mass relations, alongside the physical feature space in the ML models, may enhance the model predictions in the extrapolation region \cite{PhysRevC.106.L021301}.
\\
\\
\textit{Explainable AI -} The implementation of ML models often faces the challenge of their perceived ‘black box’ nature. To counter this issue, Explainable AI (XAI) techniques have become increasingly popular for their role in demystifying these models and enhancing understanding. Among a range of XAI techniques, SHapley Additive exPlanations (SHAP) \cite{ lundberg2017} has emerged as a prominent technique that has achieved widespread recognition.

The SHAP technique utilizes the concept of SHAP values, derived from Game Theory, which illustrates the individual contributions of players in a cooperative coalition. This concept, originally known as Shapley values \cite{shapley1953}, has been extensively studied in game theory literature \cite{winter2002}. Recently adapted to AI research, specifically in XAI, this approach treats model features as ‘players’ and the prediction as the ‘game’. SHAP values assigned to these features indicate their relative importance compared to a baseline reference. Thus, this technique effectively highlights the features most influential in the model's decision-making process.

\begin{figure}
 \centering
\includegraphics[width=\linewidth]{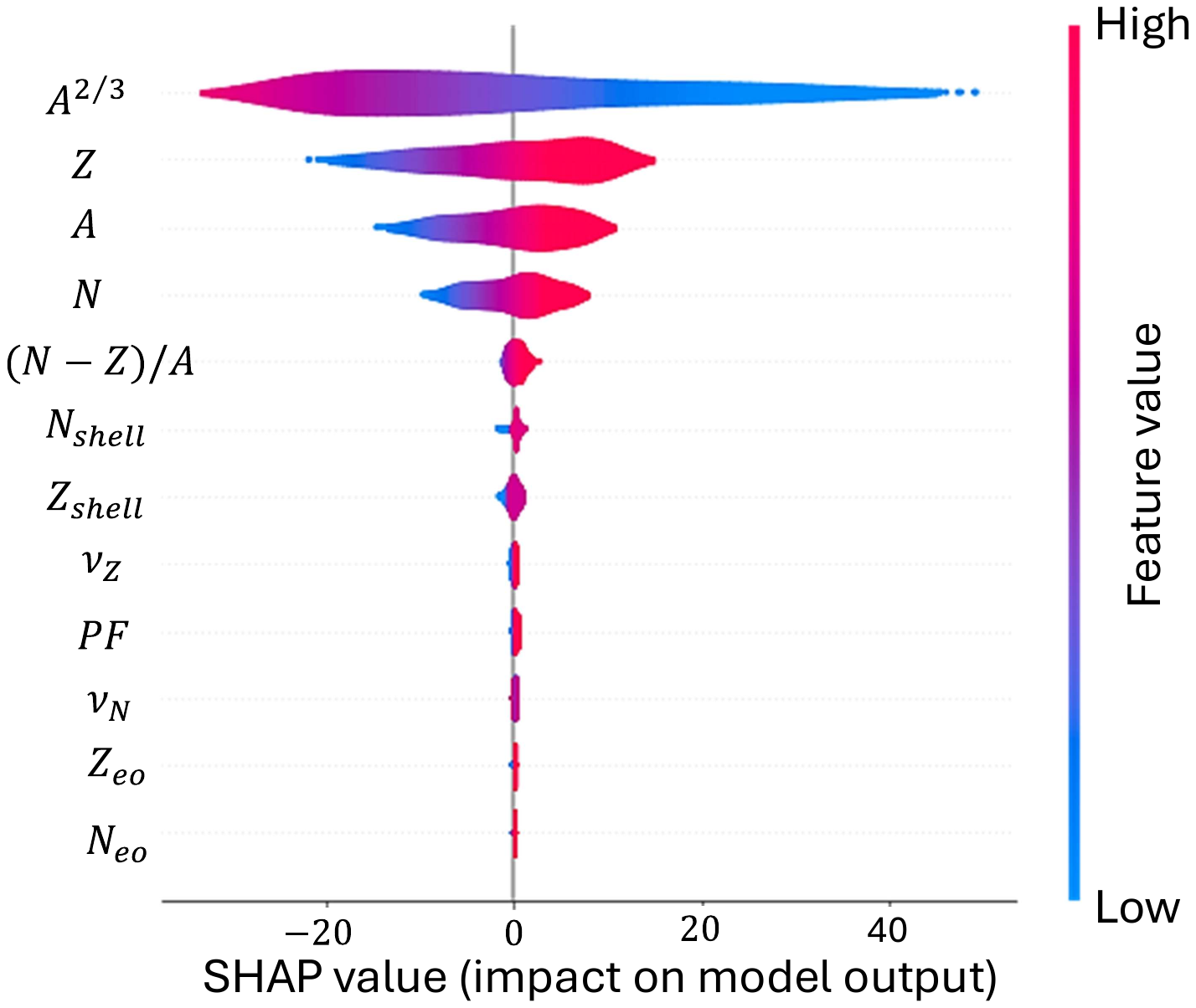}  
\caption{SHAP summary plot for the GPR-12 model. Each input is represented by a horizontal bar on the plot, where the length of the bar reflects the SHAP values’ magnitude. The color of each bar indicates the direction of the feature's influence on the prediction: blue for a decrease with lower feature values and red for an increase with higher feature values, with the intensity of the color denoting the magnitude of the feature's value.}\label{shap}
\end{figure}

We apply the SHAP technique to interpret the results of the GPR-12 model more in depth. For the test dataset, we compute the SHAP values, where each value indicates the contribution of a specific feature to the model’s prediction. These SHAP values are visually summarized in the Figure \ref{shap}. The SHAP summary plot offers an insightful illustration of how each feature influences the predictions by the GPR-12 model. In this plot, features are ordered on the y-axis based on their impact, with the most impactful feature positioned at the top and the least impactful at the bottom. To manage the extensive computational demands of calculating SHAP values for the GPR-12 model, we adhered to the guidelines suggested in the official SHAP documentation \footnote{https://shap-lrjball.readthedocs.io/en/latest/examples.html}, utilizing k-means clustering on the training data. We condensed the training data into three clusters using k-means, assigning weights to each cluster proportionate to the number of data points it encompasses. Experiments with varying numbers of clusters, including more than three, consistently yielded comparable results.

\begin{figure*} 
 \centering
 \includegraphics[width=0.9\linewidth]{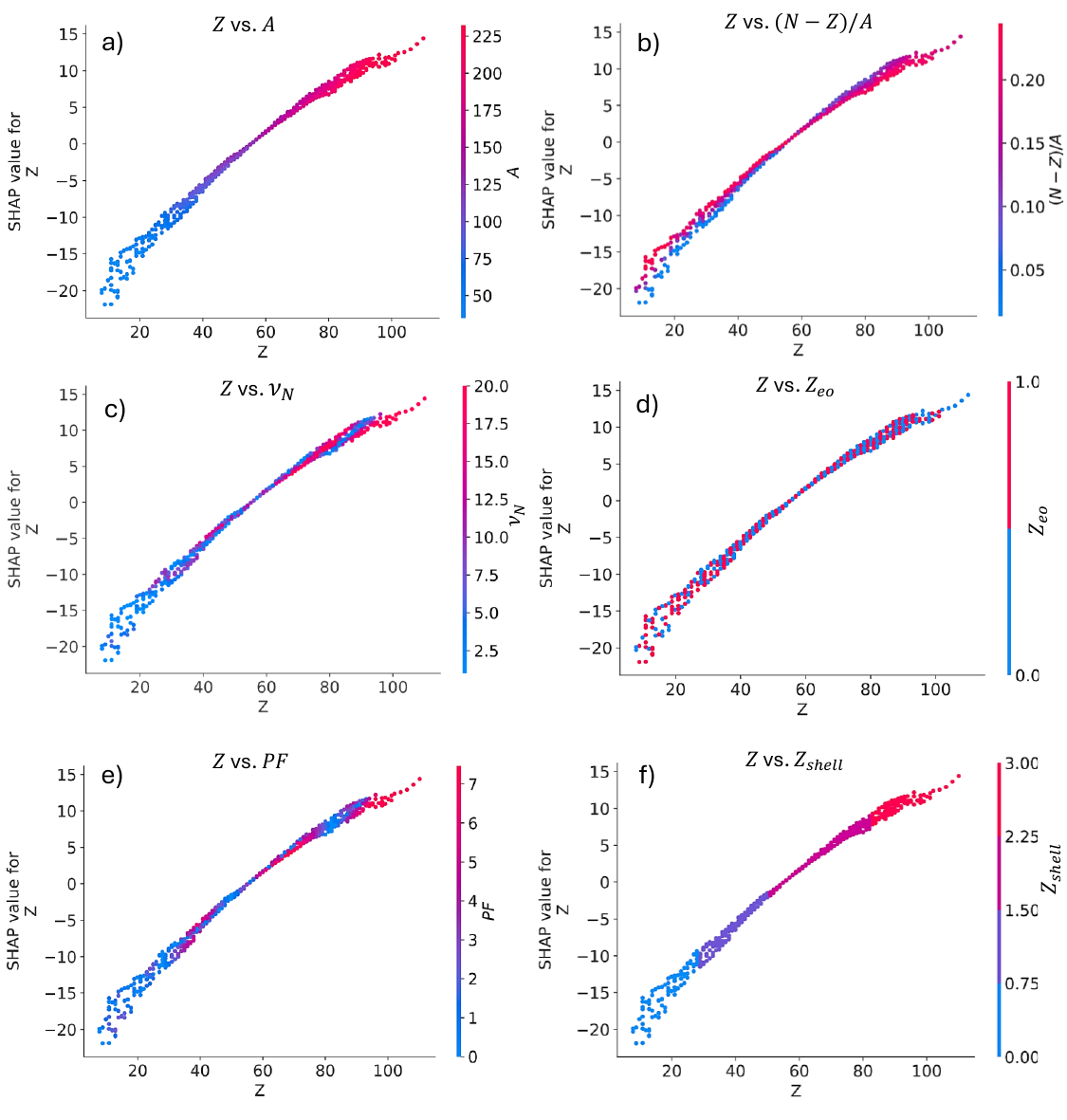} 
 \caption{Selected SHAP interaction plots. In these plots, an intense red color indicates higher positive SHAP values, while a deep blue color signifies lower negative SHAP values.}\label{shap_interactions}
\end{figure*}

The analysis reveals that $A^{2/3}$ is the most impactful factor in predicting the mass, as shown by the SHAP values. It is closely followed by $Z$, $A$ and $N$, both making noteworthy contributions to the model's predictions. In contrast, $Z_{eo}$ and $N_{eo}$ demonstrate a limited impact on predicting the mass, as indicated by their lower placement on the plot. Nonetheless, their inclusion is important to improve model predictions as we explain above in Fig.\ref{fig:3}. The SHAP values depicted in the Figure \ref{shap} clearly show the extent of each input's contribution to individual predictions. The $x$-axis represents the relative importance of each feature based on their SHAP values. Inputs with larger absolute SHAP values indicate a more significant effect on the model’s predictions, whereas those with smaller absolute values have a lesser influence. It is worth noting that we also examined the SHAP summary plot for the SVR-12 model, and the results are found to be identical. In Figure \ref{shap}, the contributions of features beyond the top five may appear minimal. However, as previously explained, the GPR-12 model outperforms its versions with fewer features (see Fig. \ref{fig:3}). This enhanced performance of the GPR-12 model can be attributed to the interactions between features, which can also be examined in detail through SHAP analysis. The SHAP analysis provides us with the opportunity to visualize the binary interactions between features. Although we can pinpoint the most impactful features in the ML models using SHAP summary plots as shown in Figure \ref{shap}, interactions between these features also play an important role in the models' performance. SHAP interaction plots provide us with an opportunity to observe the interactions of features across different parts of the nuclear chart and better understand the working mechanism of the ML models by making them more transparent.
\\
\\
In Figure \ref{shap_interactions}, we present selected interaction plots derived from the SHAP values of the GPR-12 model. While many interaction plots can be generated based on SHAP analysis, we choose to focus on interactions between the feature proton number $Z$ and others to simplify our discussion.  A majority of red in the interaction plots suggest a positive joint contribution of both features to the model's prediction. This means that higher values of these features together are likely to elevate the model's output. In contrast, a majority of blue suggests that the combined features negatively influence the model’s prediction, with lower values of both features together expected to decrease the model's output. Thus, we can pinpoint critical feature interactions and enhance our understanding of the model's decision-making based on feature combinations. For instance, the combined effect of higher values of $Z$ and $A$ (see Fig. \ref{shap_interactions}(a)) impacts the model's prediction positively, while lower values have a negative impact on the output. A similar situation is also observed among $Z$, $\nu_{N}$ (c), and $Z_{\text{shell}}$ (f). It is also seen that the interaction between $Z$ and $PF$ (e) shows variations according to the region of interest. Nonetheless, low values of $Z$ and $PF$ have a negative impact on the output. On the other hand, there is no such interaction between $Z$ and $(N-Z)/A$ or $Z_{eo}$, as shown in panels (b) and (d) of Fig. \ref{shap_interactions}. The combined effects of $Z$ and $(N-Z)/A$, and $Z$ and $Z_{eo}$ can demonstrate both positive and negative impacts across all regions. It is clear that, in the majority of plots, the interaction of proton number $Z$ with other features demonstrates a negative impact on the predictions of nuclei with low mass. Similar results are also observed for other impactful features, such as the neutron number ($N$) and mass number ($A$), indicating the necessity to identify relevant features to probe these regions more effectively. Therefore, interaction plots can be useful for identifying the relevant features to enhance predictions of ML models in regions with low prediction capability.

\section{Conclusion}
This study presents successful implementations of two ML models, SVR and GPR, using the available experimental data and physics-based feature space to make predictions for the mass excess of atomic nuclei. The ML models achieve good results not only in accurately predicting nuclear mass excesses for training and test sets but also in demonstrating robust extrapolation capabilities. Our comprehensive analysis, which includes the extrapolation region using the newly measured data from AME2020 and the region beyond, underscores the models' success in handling a diverse range of nuclear data. In addition to demonstrating the effective application of ML models, our study incorporates SHAP, an Explainable AI (XAI) technique, enhancing the interpretability of our ML models.

It is evident that SVR and GPR can be effectively utilized as reliable and efficient tools for predicting mass excess of atomic nuclei. This study highlights the potential of these ML models as powerful tools in nuclear physics and opens up new avenues for future research. These ML models can be further refined to improve their performance, especially near the drip lines. While the chosen ML models demonstrated success in predicting the mass excess of atomic nuclei, their potential applications in exploring additional nuclear properties and evaluating their performance remain as tasks for future research.
\section{Acknowledgements} 
E.Y. acknowledges support from the Science and Technology Facilities Council (UK) through Grant No. ST/Y000013/1.

\bibliographystyle{apsrev4-2}
\bibliography{bibl}

\begin{thebibliography}{54}%
\makeatletter
\providecommand \@ifxundefined [1]{%
 \@ifx{#1\undefined}
}%
\providecommand \@ifnum [1]{%
 \ifnum #1\expandafter \@firstoftwo
 \else \expandafter \@secondoftwo
 \fi
}%
\providecommand \@ifx [1]{%
 \ifx #1\expandafter \@firstoftwo
 \else \expandafter \@secondoftwo
 \fi
}%
\providecommand \natexlab [1]{#1}%
\providecommand \enquote  [1]{``#1''}%
\providecommand \bibnamefont  [1]{#1}%
\providecommand \bibfnamefont [1]{#1}%
\providecommand \citenamefont [1]{#1}%
\providecommand \href@noop [0]{\@secondoftwo}%
\providecommand \href [0]{\begingroup \@sanitize@url \@href}%
\providecommand \@href[1]{\@@startlink{#1}\@@href}%
\providecommand \@@href[1]{\endgroup#1\@@endlink}%
\providecommand \@sanitize@url [0]{\catcode `\\12\catcode `\$12\catcode
  `\&12\catcode `\#12\catcode `\^12\catcode `\_12\catcode `\%12\relax}%
\providecommand \@@startlink[1]{}%
\providecommand \@@endlink[0]{}%
\providecommand \url  [0]{\begingroup\@sanitize@url \@url }%
\providecommand \@url [1]{\endgroup\@href {#1}{\urlprefix }}%
\providecommand \urlprefix  [0]{URL }%
\providecommand \Eprint [0]{\href }%
\providecommand \doibase [0]{https://doi.org/}%
\providecommand \selectlanguage [0]{\@gobble}%
\providecommand \bibinfo  [0]{\@secondoftwo}%
\providecommand \bibfield  [0]{\@secondoftwo}%
\providecommand \translation [1]{[#1]}%
\providecommand \BibitemOpen [0]{}%
\providecommand \bibitemStop [0]{}%
\providecommand \bibitemNoStop [0]{.\EOS\space}%
\providecommand \EOS [0]{\spacefactor3000\relax}%
\providecommand \BibitemShut  [1]{\csname bibitem#1\endcsname}%
\let\auto@bib@innerbib\@empty
\bibitem [{\citenamefont {Mumpower}\ \emph {et~al.}(2015)\citenamefont
  {Mumpower}, \citenamefont {Surman}, \citenamefont {Fang}, \citenamefont
  {Beard}, \citenamefont {M\"oller}, \citenamefont {Kawano},\ and\
  \citenamefont {Aprahamian}}]{PhysRevC.92.035807}%
  \BibitemOpen
  \bibfield  {author} {\bibinfo {author} {\bibfnamefont {M.~R.}\ \bibnamefont
  {Mumpower}}, \bibinfo {author} {\bibfnamefont {R.}~\bibnamefont {Surman}},
  \bibinfo {author} {\bibfnamefont {D.-L.}\ \bibnamefont {Fang}}, \bibinfo
  {author} {\bibfnamefont {M.}~\bibnamefont {Beard}}, \bibinfo {author}
  {\bibfnamefont {P.}~\bibnamefont {M\"oller}}, \bibinfo {author}
  {\bibfnamefont {T.}~\bibnamefont {Kawano}},\ and\ \bibinfo {author}
  {\bibfnamefont {A.}~\bibnamefont {Aprahamian}},\ }\href
  {https://doi.org/10.1103/PhysRevC.92.035807} {\bibfield  {journal} {\bibinfo
  {journal} {Phys. Rev. C}\ }\textbf {\bibinfo {volume} {92}},\ \bibinfo
  {pages} {035807} (\bibinfo {year} {2015})}\BibitemShut {NoStop}%
\bibitem [{\citenamefont {Martin}\ \emph {et~al.}(2016)\citenamefont {Martin},
  \citenamefont {Arcones}, \citenamefont {Nazarewicz},\ and\ \citenamefont
  {Olsen}}]{PhysRevLett.116.121101}%
  \BibitemOpen
  \bibfield  {author} {\bibinfo {author} {\bibfnamefont {D.}~\bibnamefont
  {Martin}}, \bibinfo {author} {\bibfnamefont {A.}~\bibnamefont {Arcones}},
  \bibinfo {author} {\bibfnamefont {W.}~\bibnamefont {Nazarewicz}},\ and\
  \bibinfo {author} {\bibfnamefont {E.}~\bibnamefont {Olsen}},\ }\href
  {https://doi.org/10.1103/PhysRevLett.116.121101} {\bibfield  {journal}
  {\bibinfo  {journal} {Phys. Rev. Lett.}\ }\textbf {\bibinfo {volume} {116}},\
  \bibinfo {pages} {121101} (\bibinfo {year} {2016})}\BibitemShut {NoStop}%
\bibitem [{\citenamefont {and}\ \emph {et~al.}(2017)\citenamefont {and},
  \citenamefont {Kondev}, ,\ and\ \citenamefont {and}}]{Wang_2017}%
  \BibitemOpen
  \bibfield  {author} {\bibinfo {author} {\bibnamefont {and}}, \bibinfo
  {author} {\bibfnamefont {F.~G.}\ \bibnamefont {Kondev}}, ,\ and\ \bibinfo
  {author} {\bibfnamefont {S.~N.}\ \bibnamefont {and}},\ }\href
  {https://doi.org/10.1088/1674-1137/41/3/030003} {\bibfield  {journal}
  {\bibinfo  {journal} {Chinese Physics C}\ }\textbf {\bibinfo {volume} {41}},\
  \bibinfo {pages} {030003} (\bibinfo {year} {2017})}\BibitemShut {NoStop}%
\bibitem [{\citenamefont {Wang}\ \emph {et~al.}(2021)\citenamefont {Wang},
  \citenamefont {Huang}, \citenamefont {Kondev}, \citenamefont {Audi},\ and\
  \citenamefont {Naimi}}]{Wang_2021}%
  \BibitemOpen
  \bibfield  {author} {\bibinfo {author} {\bibfnamefont {M.}~\bibnamefont
  {Wang}}, \bibinfo {author} {\bibfnamefont {W.}~\bibnamefont {Huang}},
  \bibinfo {author} {\bibfnamefont {F.}~\bibnamefont {Kondev}}, \bibinfo
  {author} {\bibfnamefont {G.}~\bibnamefont {Audi}},\ and\ \bibinfo {author}
  {\bibfnamefont {S.}~\bibnamefont {Naimi}},\ }\href
  {https://doi.org/10.1088/1674-1137/abddaf} {\bibfield  {journal} {\bibinfo
  {journal} {Chinese Physics C}\ }\textbf {\bibinfo {volume} {45}},\ \bibinfo
  {pages} {030003} (\bibinfo {year} {2021})}\BibitemShut {NoStop}%
\bibitem [{\citenamefont {Ahn}\ \emph {et~al.}(2019)\citenamefont {Ahn},
  \citenamefont {Fukuda}, \citenamefont {Geissel}, \citenamefont {Inabe},
  \citenamefont {Iwasa}, \citenamefont {Kubo}, \citenamefont {Kusaka},
  \citenamefont {Morrissey}, \citenamefont {Murai}, \citenamefont {Nakamura},
  \citenamefont {Ohtake}, \citenamefont {Otsu}, \citenamefont {Sato},
  \citenamefont {Sherrill}, \citenamefont {Shimizu}, \citenamefont {Suzuki},
  \citenamefont {Takeda}, \citenamefont {Tarasov}, \citenamefont {Ueno},
  \citenamefont {Yanagisawa},\ and\ \citenamefont
  {Yoshida}}]{PhysRevLett.123.212501}%
  \BibitemOpen
  \bibfield  {author} {\bibinfo {author} {\bibfnamefont {D.~S.}\ \bibnamefont
  {Ahn}}, \bibinfo {author} {\bibfnamefont {N.}~\bibnamefont {Fukuda}},
  \bibinfo {author} {\bibfnamefont {H.}~\bibnamefont {Geissel}}, \bibinfo
  {author} {\bibfnamefont {N.}~\bibnamefont {Inabe}}, \bibinfo {author}
  {\bibfnamefont {N.}~\bibnamefont {Iwasa}}, \bibinfo {author} {\bibfnamefont
  {T.}~\bibnamefont {Kubo}}, \bibinfo {author} {\bibfnamefont {K.}~\bibnamefont
  {Kusaka}}, \bibinfo {author} {\bibfnamefont {D.~J.}\ \bibnamefont
  {Morrissey}}, \bibinfo {author} {\bibfnamefont {D.}~\bibnamefont {Murai}},
  \bibinfo {author} {\bibfnamefont {T.}~\bibnamefont {Nakamura}}, \bibinfo
  {author} {\bibfnamefont {M.}~\bibnamefont {Ohtake}}, \bibinfo {author}
  {\bibfnamefont {H.}~\bibnamefont {Otsu}}, \bibinfo {author} {\bibfnamefont
  {H.}~\bibnamefont {Sato}}, \bibinfo {author} {\bibfnamefont {B.~M.}\
  \bibnamefont {Sherrill}}, \bibinfo {author} {\bibfnamefont {Y.}~\bibnamefont
  {Shimizu}}, \bibinfo {author} {\bibfnamefont {H.}~\bibnamefont {Suzuki}},
  \bibinfo {author} {\bibfnamefont {H.}~\bibnamefont {Takeda}}, \bibinfo
  {author} {\bibfnamefont {O.~B.}\ \bibnamefont {Tarasov}}, \bibinfo {author}
  {\bibfnamefont {H.}~\bibnamefont {Ueno}}, \bibinfo {author} {\bibfnamefont
  {Y.}~\bibnamefont {Yanagisawa}},\ and\ \bibinfo {author} {\bibfnamefont
  {K.}~\bibnamefont {Yoshida}},\ }\href
  {https://doi.org/10.1103/PhysRevLett.123.212501} {\bibfield  {journal}
  {\bibinfo  {journal} {Phys. Rev. Lett.}\ }\textbf {\bibinfo {volume} {123}},\
  \bibinfo {pages} {212501} (\bibinfo {year} {2019})}\BibitemShut {NoStop}%
\bibitem [{\citenamefont {Wang}\ \emph {et~al.}(2014)\citenamefont {Wang},
  \citenamefont {Liu}, \citenamefont {Wu},\ and\ \citenamefont
  {Meng}}]{WANG2014215}%
  \BibitemOpen
  \bibfield  {author} {\bibinfo {author} {\bibfnamefont {N.}~\bibnamefont
  {Wang}}, \bibinfo {author} {\bibfnamefont {M.}~\bibnamefont {Liu}}, \bibinfo
  {author} {\bibfnamefont {X.}~\bibnamefont {Wu}},\ and\ \bibinfo {author}
  {\bibfnamefont {J.}~\bibnamefont {Meng}},\ }\href
  {https://doi.org/https://doi.org/10.1016/j.physletb.2014.05.049} {\bibfield
  {journal} {\bibinfo  {journal} {Physics Letters B}\ }\textbf {\bibinfo
  {volume} {734}},\ \bibinfo {pages} {215} (\bibinfo {year}
  {2014})}\BibitemShut {NoStop}%
\bibitem [{\citenamefont {Möller}\ \emph {et~al.}(2016)\citenamefont
  {Möller}, \citenamefont {Sierk}, \citenamefont {Ichikawa},\ and\
  \citenamefont {Sagawa}}]{MOLLER20161}%
  \BibitemOpen
  \bibfield  {author} {\bibinfo {author} {\bibfnamefont {P.}~\bibnamefont
  {Möller}}, \bibinfo {author} {\bibfnamefont {A.}~\bibnamefont {Sierk}},
  \bibinfo {author} {\bibfnamefont {T.}~\bibnamefont {Ichikawa}},\ and\
  \bibinfo {author} {\bibfnamefont {H.}~\bibnamefont {Sagawa}},\ }\href
  {https://doi.org/https://doi.org/10.1016/j.adt.2015.10.002} {\bibfield
  {journal} {\bibinfo  {journal} {Atomic Data and Nuclear Data Tables}\
  }\textbf {\bibinfo {volume} {109-110}},\ \bibinfo {pages} {1} (\bibinfo
  {year} {2016})}\BibitemShut {NoStop}%
\bibitem [{\citenamefont {Zhang}\ \emph {et~al.}(2022)\citenamefont {Zhang},
  \citenamefont {Cheoun}, \citenamefont {Choi}, \citenamefont {Chong},
  \citenamefont {Dong}, \citenamefont {Dong}, \citenamefont {Du}, \citenamefont
  {Geng}, \citenamefont {Ha}, \citenamefont {He}, \citenamefont {Heo},
  \citenamefont {Ho}, \citenamefont {In}, \citenamefont {Kim}, \citenamefont
  {Kim}, \citenamefont {Lee}, \citenamefont {Lee}, \citenamefont {Li},
  \citenamefont {Li}, \citenamefont {Luo}, \citenamefont {Meng}, \citenamefont
  {Mun}, \citenamefont {Niu}, \citenamefont {Pan}, \citenamefont
  {Papakonstantinou}, \citenamefont {Shang}, \citenamefont {Shen},
  \citenamefont {Shen}, \citenamefont {Sun}, \citenamefont {Sun}, \citenamefont
  {Tam}, \citenamefont {Thaivayongnou}, \citenamefont {Wang}, \citenamefont
  {Wang}, \citenamefont {Wong}, \citenamefont {Wu}, \citenamefont {Wu},
  \citenamefont {Xia}, \citenamefont {Yan}, \citenamefont {Yeung},
  \citenamefont {Yiu}, \citenamefont {Zhang}, \citenamefont {Zhang},
  \citenamefont {Zhang}, \citenamefont {Zhao},\ and\ \citenamefont
  {Zhou}}]{ZHANG2022101488}%
  \BibitemOpen
  \bibfield  {author} {\bibinfo {author} {\bibfnamefont {K.}~\bibnamefont
  {Zhang}}, \bibinfo {author} {\bibfnamefont {M.-K.}\ \bibnamefont {Cheoun}},
  \bibinfo {author} {\bibfnamefont {Y.-B.}\ \bibnamefont {Choi}}, \bibinfo
  {author} {\bibfnamefont {P.~S.}\ \bibnamefont {Chong}}, \bibinfo {author}
  {\bibfnamefont {J.}~\bibnamefont {Dong}}, \bibinfo {author} {\bibfnamefont
  {Z.}~\bibnamefont {Dong}}, \bibinfo {author} {\bibfnamefont {X.}~\bibnamefont
  {Du}}, \bibinfo {author} {\bibfnamefont {L.}~\bibnamefont {Geng}}, \bibinfo
  {author} {\bibfnamefont {E.}~\bibnamefont {Ha}}, \bibinfo {author}
  {\bibfnamefont {X.-T.}\ \bibnamefont {He}}, \bibinfo {author} {\bibfnamefont
  {C.}~\bibnamefont {Heo}}, \bibinfo {author} {\bibfnamefont {M.~C.}\
  \bibnamefont {Ho}}, \bibinfo {author} {\bibfnamefont {E.~J.}\ \bibnamefont
  {In}}, \bibinfo {author} {\bibfnamefont {S.}~\bibnamefont {Kim}}, \bibinfo
  {author} {\bibfnamefont {Y.}~\bibnamefont {Kim}}, \bibinfo {author}
  {\bibfnamefont {C.-H.}\ \bibnamefont {Lee}}, \bibinfo {author} {\bibfnamefont
  {J.}~\bibnamefont {Lee}}, \bibinfo {author} {\bibfnamefont {H.}~\bibnamefont
  {Li}}, \bibinfo {author} {\bibfnamefont {Z.}~\bibnamefont {Li}}, \bibinfo
  {author} {\bibfnamefont {T.}~\bibnamefont {Luo}}, \bibinfo {author}
  {\bibfnamefont {J.}~\bibnamefont {Meng}}, \bibinfo {author} {\bibfnamefont
  {M.-H.}\ \bibnamefont {Mun}}, \bibinfo {author} {\bibfnamefont
  {Z.}~\bibnamefont {Niu}}, \bibinfo {author} {\bibfnamefont {C.}~\bibnamefont
  {Pan}}, \bibinfo {author} {\bibfnamefont {P.}~\bibnamefont
  {Papakonstantinou}}, \bibinfo {author} {\bibfnamefont {X.}~\bibnamefont
  {Shang}}, \bibinfo {author} {\bibfnamefont {C.}~\bibnamefont {Shen}},
  \bibinfo {author} {\bibfnamefont {G.}~\bibnamefont {Shen}}, \bibinfo {author}
  {\bibfnamefont {W.}~\bibnamefont {Sun}}, \bibinfo {author} {\bibfnamefont
  {X.-X.}\ \bibnamefont {Sun}}, \bibinfo {author} {\bibfnamefont {C.~K.}\
  \bibnamefont {Tam}}, \bibinfo {author} {\bibnamefont {Thaivayongnou}},
  \bibinfo {author} {\bibfnamefont {C.}~\bibnamefont {Wang}}, \bibinfo {author}
  {\bibfnamefont {X.}~\bibnamefont {Wang}}, \bibinfo {author} {\bibfnamefont
  {S.~H.}\ \bibnamefont {Wong}}, \bibinfo {author} {\bibfnamefont
  {J.}~\bibnamefont {Wu}}, \bibinfo {author} {\bibfnamefont {X.}~\bibnamefont
  {Wu}}, \bibinfo {author} {\bibfnamefont {X.}~\bibnamefont {Xia}}, \bibinfo
  {author} {\bibfnamefont {Y.}~\bibnamefont {Yan}}, \bibinfo {author}
  {\bibfnamefont {R.~W.-Y.}\ \bibnamefont {Yeung}}, \bibinfo {author}
  {\bibfnamefont {T.~C.}\ \bibnamefont {Yiu}}, \bibinfo {author} {\bibfnamefont
  {S.}~\bibnamefont {Zhang}}, \bibinfo {author} {\bibfnamefont
  {W.}~\bibnamefont {Zhang}}, \bibinfo {author} {\bibfnamefont
  {X.}~\bibnamefont {Zhang}}, \bibinfo {author} {\bibfnamefont
  {Q.}~\bibnamefont {Zhao}},\ and\ \bibinfo {author} {\bibfnamefont {S.-G.}\
  \bibnamefont {Zhou}},\ }\href
  {https://doi.org/https://doi.org/10.1016/j.adt.2022.101488} {\bibfield
  {journal} {\bibinfo  {journal} {Atomic Data and Nuclear Data Tables}\
  }\textbf {\bibinfo {volume} {144}},\ \bibinfo {pages} {101488} (\bibinfo
  {year} {2022})}\BibitemShut {NoStop}%
\bibitem [{\citenamefont {Xia}\ \emph {et~al.}(2018)\citenamefont {Xia},
  \citenamefont {Lim}, \citenamefont {Zhao}, \citenamefont {Liang},
  \citenamefont {Qu}, \citenamefont {Chen}, \citenamefont {Liu}, \citenamefont
  {Zhang}, \citenamefont {Zhang}, \citenamefont {Kim},\ and\ \citenamefont
  {Meng}}]{XIA20181}%
  \BibitemOpen
  \bibfield  {author} {\bibinfo {author} {\bibfnamefont {X.}~\bibnamefont
  {Xia}}, \bibinfo {author} {\bibfnamefont {Y.}~\bibnamefont {Lim}}, \bibinfo
  {author} {\bibfnamefont {P.}~\bibnamefont {Zhao}}, \bibinfo {author}
  {\bibfnamefont {H.}~\bibnamefont {Liang}}, \bibinfo {author} {\bibfnamefont
  {X.}~\bibnamefont {Qu}}, \bibinfo {author} {\bibfnamefont {Y.}~\bibnamefont
  {Chen}}, \bibinfo {author} {\bibfnamefont {H.}~\bibnamefont {Liu}}, \bibinfo
  {author} {\bibfnamefont {L.}~\bibnamefont {Zhang}}, \bibinfo {author}
  {\bibfnamefont {S.}~\bibnamefont {Zhang}}, \bibinfo {author} {\bibfnamefont
  {Y.}~\bibnamefont {Kim}},\ and\ \bibinfo {author} {\bibfnamefont
  {J.}~\bibnamefont {Meng}},\ }\href
  {https://doi.org/https://doi.org/10.1016/j.adt.2017.09.001} {\bibfield
  {journal} {\bibinfo  {journal} {Atomic Data and Nuclear Data Tables}\
  }\textbf {\bibinfo {volume} {121-122}},\ \bibinfo {pages} {1} (\bibinfo
  {year} {2018})}\BibitemShut {NoStop}%
\bibitem [{\citenamefont {Afanasjev}\ and\ \citenamefont
  {Agbemava}(2016)}]{PhysRevC.93.054310}%
  \BibitemOpen
  \bibfield  {author} {\bibinfo {author} {\bibfnamefont {A.~V.}\ \bibnamefont
  {Afanasjev}}\ and\ \bibinfo {author} {\bibfnamefont {S.~E.}\ \bibnamefont
  {Agbemava}},\ }\href {https://doi.org/10.1103/PhysRevC.93.054310} {\bibfield
  {journal} {\bibinfo  {journal} {Phys. Rev. C}\ }\textbf {\bibinfo {volume}
  {93}},\ \bibinfo {pages} {054310} (\bibinfo {year} {2016})}\BibitemShut
  {NoStop}%
\bibitem [{\citenamefont {Afanasjev}\ \emph {et~al.}(2015)\citenamefont
  {Afanasjev}, \citenamefont {Agbemava}, \citenamefont {Ray},\ and\
  \citenamefont {Ring}}]{PhysRevC.91.014324}%
  \BibitemOpen
  \bibfield  {author} {\bibinfo {author} {\bibfnamefont {A.~V.}\ \bibnamefont
  {Afanasjev}}, \bibinfo {author} {\bibfnamefont {S.~E.}\ \bibnamefont
  {Agbemava}}, \bibinfo {author} {\bibfnamefont {D.}~\bibnamefont {Ray}},\ and\
  \bibinfo {author} {\bibfnamefont {P.}~\bibnamefont {Ring}},\ }\href
  {https://doi.org/10.1103/PhysRevC.91.014324} {\bibfield  {journal} {\bibinfo
  {journal} {Phys. Rev. C}\ }\textbf {\bibinfo {volume} {91}},\ \bibinfo
  {pages} {014324} (\bibinfo {year} {2015})}\BibitemShut {NoStop}%
\bibitem [{\citenamefont {Agbemava}\ \emph {et~al.}(2014)\citenamefont
  {Agbemava}, \citenamefont {Afanasjev}, \citenamefont {Ray},\ and\
  \citenamefont {Ring}}]{PhysRevC.89.054320}%
  \BibitemOpen
  \bibfield  {author} {\bibinfo {author} {\bibfnamefont {S.~E.}\ \bibnamefont
  {Agbemava}}, \bibinfo {author} {\bibfnamefont {A.~V.}\ \bibnamefont
  {Afanasjev}}, \bibinfo {author} {\bibfnamefont {D.}~\bibnamefont {Ray}},\
  and\ \bibinfo {author} {\bibfnamefont {P.}~\bibnamefont {Ring}},\ }\href
  {https://doi.org/10.1103/PhysRevC.89.054320} {\bibfield  {journal} {\bibinfo
  {journal} {Phys. Rev. C}\ }\textbf {\bibinfo {volume} {89}},\ \bibinfo
  {pages} {054320} (\bibinfo {year} {2014})}\BibitemShut {NoStop}%
\bibitem [{\citenamefont {Erler}\ \emph {et~al.}(2012)\citenamefont {Erler},
  \citenamefont {Birge}, \citenamefont {Kortelainen}, \citenamefont
  {Nazarewicz}, \citenamefont {Olsen}, \citenamefont {Perhac},\ and\
  \citenamefont {Stoitsov}}]{Erler2012}%
  \BibitemOpen
  \bibfield  {author} {\bibinfo {author} {\bibfnamefont {J.}~\bibnamefont
  {Erler}}, \bibinfo {author} {\bibfnamefont {N.}~\bibnamefont {Birge}},
  \bibinfo {author} {\bibfnamefont {M.}~\bibnamefont {Kortelainen}}, \bibinfo
  {author} {\bibfnamefont {W.}~\bibnamefont {Nazarewicz}}, \bibinfo {author}
  {\bibfnamefont {E.}~\bibnamefont {Olsen}}, \bibinfo {author} {\bibfnamefont
  {A.~M.}\ \bibnamefont {Perhac}},\ and\ \bibinfo {author} {\bibfnamefont
  {M.}~\bibnamefont {Stoitsov}},\ }\href {https://doi.org/10.1038/nature11188}
  {\bibfield  {journal} {\bibinfo  {journal} {Nature}\ }\textbf {\bibinfo
  {volume} {486}},\ \bibinfo {pages} {509} (\bibinfo {year}
  {2012})}\BibitemShut {NoStop}%
\bibitem [{\citenamefont {Grams}\ \emph {et~al.}(2023)\citenamefont {Grams},
  \citenamefont {Ryssens}, \citenamefont {Scamps}, \citenamefont {Goriely},\
  and\ \citenamefont {Chamel}}]{Grams2023}%
  \BibitemOpen
  \bibfield  {author} {\bibinfo {author} {\bibfnamefont {G.}~\bibnamefont
  {Grams}}, \bibinfo {author} {\bibfnamefont {W.}~\bibnamefont {Ryssens}},
  \bibinfo {author} {\bibfnamefont {G.}~\bibnamefont {Scamps}}, \bibinfo
  {author} {\bibfnamefont {S.}~\bibnamefont {Goriely}},\ and\ \bibinfo {author}
  {\bibfnamefont {N.}~\bibnamefont {Chamel}},\ }\href
  {https://doi.org/10.1140/epja/s10050-023-01158-6} {\bibfield  {journal}
  {\bibinfo  {journal} {The European Physical Journal A}\ }\textbf {\bibinfo
  {volume} {59}},\ \bibinfo {pages} {270} (\bibinfo {year} {2023})}\BibitemShut
  {NoStop}%
\bibitem [{\citenamefont {Ravli\ifmmode~\acute{c}\else \'{c}\fi{}}\ \emph
  {et~al.}(2023)\citenamefont {Ravli\ifmmode~\acute{c}\else \'{c}\fi{}},
  \citenamefont {Y\"uksel}, \citenamefont {Nik\ifmmode \check{s}\else
  \v{s}\fi{}i\ifmmode~\acute{c}\else \'{c}\fi{}},\ and\ \citenamefont
  {Paar}}]{PhysRevC.108.054305}%
  \BibitemOpen
  \bibfield  {author} {\bibinfo {author} {\bibfnamefont {A.}~\bibnamefont
  {Ravli\ifmmode~\acute{c}\else \'{c}\fi{}}}, \bibinfo {author} {\bibfnamefont
  {E.}~\bibnamefont {Y\"uksel}}, \bibinfo {author} {\bibfnamefont
  {T.}~\bibnamefont {Nik\ifmmode \check{s}\else
  \v{s}\fi{}i\ifmmode~\acute{c}\else \'{c}\fi{}}},\ and\ \bibinfo {author}
  {\bibfnamefont {N.}~\bibnamefont {Paar}},\ }\href
  {https://doi.org/10.1103/PhysRevC.108.054305} {\bibfield  {journal} {\bibinfo
   {journal} {Phys. Rev. C}\ }\textbf {\bibinfo {volume} {108}},\ \bibinfo
  {pages} {054305} (\bibinfo {year} {2023})}\BibitemShut {NoStop}%
\bibitem [{\citenamefont {Ravli\ifmmode~\acute{c}\else \'{c}\fi{}}\ \emph
  {et~al.}(2024)\citenamefont {Ravli\ifmmode~\acute{c}\else \'{c}\fi{}},
  \citenamefont {Y\"uksel}, \citenamefont {Nik\ifmmode \check{s}\else
  \v{s}\fi{}i\ifmmode~\acute{c}\else \'{c}\fi{}},\ and\ \citenamefont
  {Paar}}]{ravlic2023global}%
  \BibitemOpen
  \bibfield  {author} {\bibinfo {author} {\bibfnamefont {A.}~\bibnamefont
  {Ravli\ifmmode~\acute{c}\else \'{c}\fi{}}}, \bibinfo {author} {\bibfnamefont
  {E.}~\bibnamefont {Y\"uksel}}, \bibinfo {author} {\bibfnamefont
  {T.}~\bibnamefont {Nik\ifmmode \check{s}\else
  \v{s}\fi{}i\ifmmode~\acute{c}\else \'{c}\fi{}}},\ and\ \bibinfo {author}
  {\bibfnamefont {N.}~\bibnamefont {Paar}},\ }\href
  {https://doi.org/10.1103/PhysRevC.109.014318} {\bibfield  {journal} {\bibinfo
   {journal} {Phys. Rev. C}\ }\textbf {\bibinfo {volume} {109}},\ \bibinfo
  {pages} {014318} (\bibinfo {year} {2024})}\BibitemShut {NoStop}%
\bibitem [{\citenamefont {Ravli{\'{c}}}\ \emph {et~al.}(2023)\citenamefont
  {Ravli{\'{c}}}, \citenamefont {Y{\"u}ksel}, \citenamefont
  {Nik{\v{s}}i{\'{c}}},\ and\ \citenamefont {Paar}}]{Ravlic2023}%
  \BibitemOpen
  \bibfield  {author} {\bibinfo {author} {\bibfnamefont {A.}~\bibnamefont
  {Ravli{\'{c}}}}, \bibinfo {author} {\bibfnamefont {E.}~\bibnamefont
  {Y{\"u}ksel}}, \bibinfo {author} {\bibfnamefont {T.}~\bibnamefont
  {Nik{\v{s}}i{\'{c}}}},\ and\ \bibinfo {author} {\bibfnamefont
  {N.}~\bibnamefont {Paar}},\ }\href
  {https://doi.org/10.1038/s41467-023-40613-2} {\bibfield  {journal} {\bibinfo
  {journal} {Nature Communications}\ }\textbf {\bibinfo {volume} {14}},\
  \bibinfo {pages} {4834} (\bibinfo {year} {2023})}\BibitemShut {NoStop}%
\bibitem [{\citenamefont {Goriely}\ \emph
  {et~al.}(2013{\natexlab{a}})\citenamefont {Goriely}, \citenamefont {Chamel},\
  and\ \citenamefont {Pearson}}]{PhysRevC.88.024308}%
  \BibitemOpen
  \bibfield  {author} {\bibinfo {author} {\bibfnamefont {S.}~\bibnamefont
  {Goriely}}, \bibinfo {author} {\bibfnamefont {N.}~\bibnamefont {Chamel}},\
  and\ \bibinfo {author} {\bibfnamefont {J.~M.}\ \bibnamefont {Pearson}},\
  }\href {https://doi.org/10.1103/PhysRevC.88.024308} {\bibfield  {journal}
  {\bibinfo  {journal} {Phys. Rev. C}\ }\textbf {\bibinfo {volume} {88}},\
  \bibinfo {pages} {024308} (\bibinfo {year} {2013}{\natexlab{a}})}\BibitemShut
  {NoStop}%
\bibitem [{\citenamefont {Utama}\ and\ \citenamefont
  {Piekarewicz}(2017)}]{PhysRevC.96.044308}%
  \BibitemOpen
  \bibfield  {author} {\bibinfo {author} {\bibfnamefont {R.}~\bibnamefont
  {Utama}}\ and\ \bibinfo {author} {\bibfnamefont {J.}~\bibnamefont
  {Piekarewicz}},\ }\href {https://doi.org/10.1103/PhysRevC.96.044308}
  {\bibfield  {journal} {\bibinfo  {journal} {Phys. Rev. C}\ }\textbf {\bibinfo
  {volume} {96}},\ \bibinfo {pages} {044308} (\bibinfo {year}
  {2017})}\BibitemShut {NoStop}%
\bibitem [{\citenamefont {Goriely}\ \emph
  {et~al.}(2013{\natexlab{b}})\citenamefont {Goriely}, \citenamefont {Chamel},\
  and\ \citenamefont {Pearson}}]{PhysRevC.88.061302}%
  \BibitemOpen
  \bibfield  {author} {\bibinfo {author} {\bibfnamefont {S.}~\bibnamefont
  {Goriely}}, \bibinfo {author} {\bibfnamefont {N.}~\bibnamefont {Chamel}},\
  and\ \bibinfo {author} {\bibfnamefont {J.~M.}\ \bibnamefont {Pearson}},\
  }\href {https://doi.org/10.1103/PhysRevC.88.061302} {\bibfield  {journal}
  {\bibinfo  {journal} {Phys. Rev. C}\ }\textbf {\bibinfo {volume} {88}},\
  \bibinfo {pages} {061302} (\bibinfo {year} {2013}{\natexlab{b}})}\BibitemShut
  {NoStop}%
\bibitem [{\citenamefont {Goriely}\ \emph {et~al.}(2016)\citenamefont
  {Goriely}, \citenamefont {Chamel},\ and\ \citenamefont
  {Pearson}}]{PhysRevC.93.034337}%
  \BibitemOpen
  \bibfield  {author} {\bibinfo {author} {\bibfnamefont {S.}~\bibnamefont
  {Goriely}}, \bibinfo {author} {\bibfnamefont {N.}~\bibnamefont {Chamel}},\
  and\ \bibinfo {author} {\bibfnamefont {J.~M.}\ \bibnamefont {Pearson}},\
  }\href {https://doi.org/10.1103/PhysRevC.93.034337} {\bibfield  {journal}
  {\bibinfo  {journal} {Phys. Rev. C}\ }\textbf {\bibinfo {volume} {93}},\
  \bibinfo {pages} {034337} (\bibinfo {year} {2016})}\BibitemShut {NoStop}%
\bibitem [{\citenamefont {Boehnlein}\ \emph {et~al.}(2022)\citenamefont
  {Boehnlein}, \citenamefont {Diefenthaler}, \citenamefont {Sato},
  \citenamefont {Schram}, \citenamefont {Ziegler}, \citenamefont {Fanelli},
  \citenamefont {Hjorth-Jensen}, \citenamefont {Horn}, \citenamefont {Kuchera},
  \citenamefont {Lee}, \citenamefont {Nazarewicz}, \citenamefont {Ostroumov},
  \citenamefont {Orginos}, \citenamefont {Poon}, \citenamefont {Wang},
  \citenamefont {Scheinker}, \citenamefont {Smith},\ and\ \citenamefont
  {Pang}}]{RevModPhys.94.031003}%
  \BibitemOpen
  \bibfield  {author} {\bibinfo {author} {\bibfnamefont {A.}~\bibnamefont
  {Boehnlein}}, \bibinfo {author} {\bibfnamefont {M.}~\bibnamefont
  {Diefenthaler}}, \bibinfo {author} {\bibfnamefont {N.}~\bibnamefont {Sato}},
  \bibinfo {author} {\bibfnamefont {M.}~\bibnamefont {Schram}}, \bibinfo
  {author} {\bibfnamefont {V.}~\bibnamefont {Ziegler}}, \bibinfo {author}
  {\bibfnamefont {C.}~\bibnamefont {Fanelli}}, \bibinfo {author} {\bibfnamefont
  {M.}~\bibnamefont {Hjorth-Jensen}}, \bibinfo {author} {\bibfnamefont
  {T.}~\bibnamefont {Horn}}, \bibinfo {author} {\bibfnamefont {M.~P.}\
  \bibnamefont {Kuchera}}, \bibinfo {author} {\bibfnamefont {D.}~\bibnamefont
  {Lee}}, \bibinfo {author} {\bibfnamefont {W.}~\bibnamefont {Nazarewicz}},
  \bibinfo {author} {\bibfnamefont {P.}~\bibnamefont {Ostroumov}}, \bibinfo
  {author} {\bibfnamefont {K.}~\bibnamefont {Orginos}}, \bibinfo {author}
  {\bibfnamefont {A.}~\bibnamefont {Poon}}, \bibinfo {author} {\bibfnamefont
  {X.-N.}\ \bibnamefont {Wang}}, \bibinfo {author} {\bibfnamefont
  {A.}~\bibnamefont {Scheinker}}, \bibinfo {author} {\bibfnamefont {M.~S.}\
  \bibnamefont {Smith}},\ and\ \bibinfo {author} {\bibfnamefont {L.-G.}\
  \bibnamefont {Pang}},\ }\href {https://doi.org/10.1103/RevModPhys.94.031003}
  {\bibfield  {journal} {\bibinfo  {journal} {Rev. Mod. Phys.}\ }\textbf
  {\bibinfo {volume} {94}},\ \bibinfo {pages} {031003} (\bibinfo {year}
  {2022})}\BibitemShut {NoStop}%
\bibitem [{\citenamefont {Gazula}\ \emph {et~al.}(1992)\citenamefont {Gazula},
  \citenamefont {Clark},\ and\ \citenamefont {Bohr}}]{GAZULA19921}%
  \BibitemOpen
  \bibfield  {author} {\bibinfo {author} {\bibfnamefont {S.}~\bibnamefont
  {Gazula}}, \bibinfo {author} {\bibfnamefont {J.}~\bibnamefont {Clark}},\ and\
  \bibinfo {author} {\bibfnamefont {H.}~\bibnamefont {Bohr}},\ }\href
  {https://doi.org/https://doi.org/10.1016/0375-9474(92)90191-L} {\bibfield
  {journal} {\bibinfo  {journal} {Nuclear Physics A}\ }\textbf {\bibinfo
  {volume} {540}},\ \bibinfo {pages} {1} (\bibinfo {year} {1992})}\BibitemShut
  {NoStop}%
\bibitem [{\citenamefont {Clark}\ and\ \citenamefont {Li}(2006)}]{CLARK_2006}%
  \BibitemOpen
  \bibfield  {author} {\bibinfo {author} {\bibfnamefont {J.~W.}\ \bibnamefont
  {Clark}}\ and\ \bibinfo {author} {\bibfnamefont {H.}~\bibnamefont {Li}},\
  }\href {https://doi.org/10.1142/s0217979206036053} {\bibfield  {journal}
  {\bibinfo  {journal} {International Journal of Modern Physics B}\ }\textbf
  {\bibinfo {volume} {20}},\ \bibinfo {pages} {5015–5029} (\bibinfo {year}
  {2006})}\BibitemShut {NoStop}%
\bibitem [{\citenamefont {Athanassopoulos}\ \emph {et~al.}(2004)\citenamefont
  {Athanassopoulos}, \citenamefont {Mavrommatis}, \citenamefont {Gernoth},\
  and\ \citenamefont {Clark}}]{ATHANASSOPOULOS2004222}%
  \BibitemOpen
  \bibfield  {author} {\bibinfo {author} {\bibfnamefont {S.}~\bibnamefont
  {Athanassopoulos}}, \bibinfo {author} {\bibfnamefont {E.}~\bibnamefont
  {Mavrommatis}}, \bibinfo {author} {\bibfnamefont {K.}~\bibnamefont
  {Gernoth}},\ and\ \bibinfo {author} {\bibfnamefont {J.}~\bibnamefont
  {Clark}},\ }\href
  {https://doi.org/https://doi.org/10.1016/j.nuclphysa.2004.08.006} {\bibfield
  {journal} {\bibinfo  {journal} {Nuclear Physics A}\ }\textbf {\bibinfo
  {volume} {743}},\ \bibinfo {pages} {222} (\bibinfo {year}
  {2004})}\BibitemShut {NoStop}%
\bibitem [{\citenamefont {Y{\"u}ksel}\ \emph {et~al.}(2021)\citenamefont
  {Y{\"u}ksel}, \citenamefont {Soydaner},\ and\ \citenamefont
  {Bahtiyar}}]{yuksel2021}%
  \BibitemOpen
  \bibfield  {author} {\bibinfo {author} {\bibfnamefont {E.}~\bibnamefont
  {Y{\"u}ksel}}, \bibinfo {author} {\bibfnamefont {D.}~\bibnamefont
  {Soydaner}},\ and\ \bibinfo {author} {\bibfnamefont {H.}~\bibnamefont
  {Bahtiyar}},\ }\href {https://doi.org/10.1142/s0218301321500178} {\bibfield
  {journal} {\bibinfo  {journal} {International Journal of Modern Physics E}\
  }\textbf {\bibinfo {volume} {30}},\ \bibinfo {pages} {2150017} (\bibinfo
  {year} {2021})}\BibitemShut {NoStop}%
\bibitem [{\citenamefont {Bahtiyar}\ \emph {et~al.}(2022)\citenamefont
  {Bahtiyar}, \citenamefont {Derya},\ and\ \citenamefont
  {Yüksel}}]{BAHTIYAR2022109470}%
  \BibitemOpen
  \bibfield  {author} {\bibinfo {author} {\bibfnamefont {H.}~\bibnamefont
  {Bahtiyar}}, \bibinfo {author} {\bibnamefont {Derya}},\ and\ \bibinfo
  {author} {\bibfnamefont {E.}~\bibnamefont {Yüksel}},\ }\href
  {https://doi.org/https://doi.org/10.1016/j.asoc.2022.109470} {\bibfield
  {journal} {\bibinfo  {journal} {Applied Soft Computing}\ }\textbf {\bibinfo
  {volume} {128}},\ \bibinfo {pages} {109470} (\bibinfo {year}
  {2022})}\BibitemShut {NoStop}%
\bibitem [{\citenamefont {Li}\ \emph {et~al.}(2022)\citenamefont {Li},
  \citenamefont {Tong}, \citenamefont {Du},\ and\ \citenamefont
  {Pang}}]{PhysRevC.105.064306}%
  \BibitemOpen
  \bibfield  {author} {\bibinfo {author} {\bibfnamefont {C.-Q.}\ \bibnamefont
  {Li}}, \bibinfo {author} {\bibfnamefont {C.-N.}\ \bibnamefont {Tong}},
  \bibinfo {author} {\bibfnamefont {H.-J.}\ \bibnamefont {Du}},\ and\ \bibinfo
  {author} {\bibfnamefont {L.-G.}\ \bibnamefont {Pang}},\ }\href
  {https://doi.org/10.1103/PhysRevC.105.064306} {\bibfield  {journal} {\bibinfo
   {journal} {Phys. Rev. C}\ }\textbf {\bibinfo {volume} {105}},\ \bibinfo
  {pages} {064306} (\bibinfo {year} {2022})}\BibitemShut {NoStop}%
\bibitem [{\citenamefont {Shelley}\ and\ \citenamefont
  {Pastore}(2021)}]{universe7050131}%
  \BibitemOpen
  \bibfield  {author} {\bibinfo {author} {\bibfnamefont {M.}~\bibnamefont
  {Shelley}}\ and\ \bibinfo {author} {\bibfnamefont {A.}~\bibnamefont
  {Pastore}},\ }\bibfield  {journal} {\bibinfo  {journal} {Universe}\ }\textbf
  {\bibinfo {volume} {7}},\ \href {https://doi.org/10.3390/universe7050131}
  {10.3390/universe7050131} (\bibinfo {year} {2021})\BibitemShut {NoStop}%
\bibitem [{\citenamefont {Lovell}\ \emph {et~al.}(2022)\citenamefont {Lovell},
  \citenamefont {Mohan}, \citenamefont {Sprouse},\ and\ \citenamefont
  {Mumpower}}]{PhysRevC.106.014305}%
  \BibitemOpen
  \bibfield  {author} {\bibinfo {author} {\bibfnamefont {A.~E.}\ \bibnamefont
  {Lovell}}, \bibinfo {author} {\bibfnamefont {A.~T.}\ \bibnamefont {Mohan}},
  \bibinfo {author} {\bibfnamefont {T.~M.}\ \bibnamefont {Sprouse}},\ and\
  \bibinfo {author} {\bibfnamefont {M.~R.}\ \bibnamefont {Mumpower}},\ }\href
  {https://doi.org/10.1103/PhysRevC.106.014305} {\bibfield  {journal} {\bibinfo
   {journal} {Phys. Rev. C}\ }\textbf {\bibinfo {volume} {106}},\ \bibinfo
  {pages} {014305} (\bibinfo {year} {2022})}\BibitemShut {NoStop}%
\bibitem [{\citenamefont {Mumpower}\ \emph {et~al.}(2022)\citenamefont
  {Mumpower}, \citenamefont {Sprouse}, \citenamefont {Lovell},\ and\
  \citenamefont {Mohan}}]{PhysRevC.106.L021301}%
  \BibitemOpen
  \bibfield  {author} {\bibinfo {author} {\bibfnamefont {M.~R.}\ \bibnamefont
  {Mumpower}}, \bibinfo {author} {\bibfnamefont {T.~M.}\ \bibnamefont
  {Sprouse}}, \bibinfo {author} {\bibfnamefont {A.~E.}\ \bibnamefont
  {Lovell}},\ and\ \bibinfo {author} {\bibfnamefont {A.~T.}\ \bibnamefont
  {Mohan}},\ }\href {https://doi.org/10.1103/PhysRevC.106.L021301} {\bibfield
  {journal} {\bibinfo  {journal} {Phys. Rev. C}\ }\textbf {\bibinfo {volume}
  {106}},\ \bibinfo {pages} {L021301} (\bibinfo {year} {2022})}\BibitemShut
  {NoStop}%
\bibitem [{\citenamefont {Mumpower}\ \emph {et~al.}(2023)\citenamefont
  {Mumpower}, \citenamefont {Li}, \citenamefont {Sprouse}, \citenamefont
  {Meyer}, \citenamefont {Lovell},\ and\ \citenamefont
  {Mohan}}]{10.3389/fphy.2023.1198572}%
  \BibitemOpen
  \bibfield  {author} {\bibinfo {author} {\bibfnamefont {M.}~\bibnamefont
  {Mumpower}}, \bibinfo {author} {\bibfnamefont {M.}~\bibnamefont {Li}},
  \bibinfo {author} {\bibfnamefont {T.~M.}\ \bibnamefont {Sprouse}}, \bibinfo
  {author} {\bibfnamefont {B.~S.}\ \bibnamefont {Meyer}}, \bibinfo {author}
  {\bibfnamefont {A.~E.}\ \bibnamefont {Lovell}},\ and\ \bibinfo {author}
  {\bibfnamefont {A.~T.}\ \bibnamefont {Mohan}},\ }\bibfield  {journal}
  {\bibinfo  {journal} {Frontiers in Physics}\ }\textbf {\bibinfo {volume}
  {11}},\ \href {https://doi.org/10.3389/fphy.2023.1198572}
  {10.3389/fphy.2023.1198572} (\bibinfo {year} {2023})\BibitemShut {NoStop}%
\bibitem [{\citenamefont {Utama}\ and\ \citenamefont
  {Piekarewicz}(2018)}]{PhysRevC.97.014306}%
  \BibitemOpen
  \bibfield  {author} {\bibinfo {author} {\bibfnamefont {R.}~\bibnamefont
  {Utama}}\ and\ \bibinfo {author} {\bibfnamefont {J.}~\bibnamefont
  {Piekarewicz}},\ }\href {https://doi.org/10.1103/PhysRevC.97.014306}
  {\bibfield  {journal} {\bibinfo  {journal} {Phys. Rev. C}\ }\textbf {\bibinfo
  {volume} {97}},\ \bibinfo {pages} {014306} (\bibinfo {year}
  {2018})}\BibitemShut {NoStop}%
\bibitem [{\citenamefont {Niu}\ and\ \citenamefont {Liang}(2018)}]{NIU201848}%
  \BibitemOpen
  \bibfield  {author} {\bibinfo {author} {\bibfnamefont {Z.}~\bibnamefont
  {Niu}}\ and\ \bibinfo {author} {\bibfnamefont {H.}~\bibnamefont {Liang}},\
  }\href {https://doi.org/https://doi.org/10.1016/j.physletb.2018.01.002}
  {\bibfield  {journal} {\bibinfo  {journal} {Physics Letters B}\ }\textbf
  {\bibinfo {volume} {778}},\ \bibinfo {pages} {48} (\bibinfo {year}
  {2018})}\BibitemShut {NoStop}%
\bibitem [{\citenamefont {Neufcourt}\ \emph {et~al.}(2018)\citenamefont
  {Neufcourt}, \citenamefont {Cao}, \citenamefont {Nazarewicz},\ and\
  \citenamefont {Viens}}]{PhysRevC.98.034318}%
  \BibitemOpen
  \bibfield  {author} {\bibinfo {author} {\bibfnamefont {L.}~\bibnamefont
  {Neufcourt}}, \bibinfo {author} {\bibfnamefont {Y.}~\bibnamefont {Cao}},
  \bibinfo {author} {\bibfnamefont {W.}~\bibnamefont {Nazarewicz}},\ and\
  \bibinfo {author} {\bibfnamefont {F.}~\bibnamefont {Viens}},\ }\href
  {https://doi.org/10.1103/PhysRevC.98.034318} {\bibfield  {journal} {\bibinfo
  {journal} {Phys. Rev. C}\ }\textbf {\bibinfo {volume} {98}},\ \bibinfo
  {pages} {034318} (\bibinfo {year} {2018})}\BibitemShut {NoStop}%
\bibitem [{\citenamefont {Neufcourt}\ \emph {et~al.}(2019)\citenamefont
  {Neufcourt}, \citenamefont {Cao}, \citenamefont {Nazarewicz}, \citenamefont
  {Olsen},\ and\ \citenamefont {Viens}}]{PhysRevLett.122.062502}%
  \BibitemOpen
  \bibfield  {author} {\bibinfo {author} {\bibfnamefont {L.}~\bibnamefont
  {Neufcourt}}, \bibinfo {author} {\bibfnamefont {Y.}~\bibnamefont {Cao}},
  \bibinfo {author} {\bibfnamefont {W.}~\bibnamefont {Nazarewicz}}, \bibinfo
  {author} {\bibfnamefont {E.}~\bibnamefont {Olsen}},\ and\ \bibinfo {author}
  {\bibfnamefont {F.}~\bibnamefont {Viens}},\ }\href
  {https://doi.org/10.1103/PhysRevLett.122.062502} {\bibfield  {journal}
  {\bibinfo  {journal} {Phys. Rev. Lett.}\ }\textbf {\bibinfo {volume} {122}},\
  \bibinfo {pages} {062502} (\bibinfo {year} {2019})}\BibitemShut {NoStop}%
\bibitem [{\citenamefont {Drucker}\ \emph {et~al.}(1996)\citenamefont
  {Drucker}, \citenamefont {Burges}, \citenamefont {Kaufman}, \citenamefont
  {Smola},\ and\ \citenamefont {Vapnik}}]{drucker1996}%
  \BibitemOpen
  \bibfield  {author} {\bibinfo {author} {\bibfnamefont {H.}~\bibnamefont
  {Drucker}}, \bibinfo {author} {\bibfnamefont {C.}~\bibnamefont {Burges}},
  \bibinfo {author} {\bibfnamefont {L.}~\bibnamefont {Kaufman}}, \bibinfo
  {author} {\bibfnamefont {A.}~\bibnamefont {Smola}},\ and\ \bibinfo {author}
  {\bibfnamefont {V.}~\bibnamefont {Vapnik}},\ }\href
  {https://proceedings.neurips.cc/paper_files/paper/1996/file/d38901788c533e8286cb6400b40b386d-Paper.pdf}
  {\bibfield  {journal} {\bibinfo  {journal} {Advances in Neural Information
  Processing Systems}\ }\textbf {\bibinfo {volume} {9}} (\bibinfo {year}
  {1996})}\BibitemShut {NoStop}%
\bibitem [{\citenamefont {Boser}\ \emph {et~al.}(1992)\citenamefont {Boser},
  \citenamefont {Guyon},\ and\ \citenamefont {Vapnik}}]{boser1992}%
  \BibitemOpen
  \bibfield  {author} {\bibinfo {author} {\bibfnamefont {B.}~\bibnamefont
  {Boser}}, \bibinfo {author} {\bibfnamefont {I.}~\bibnamefont {Guyon}},\ and\
  \bibinfo {author} {\bibfnamefont {V.}~\bibnamefont {Vapnik}},\ }\href
  {https://dl.acm.org/doi/abs/10.1145/130385.130401} {\bibfield  {journal}
  {\bibinfo  {journal} {Proceedings of the Fifth Annual Workshop on
  Computational Learning Theory}\ ,\ \bibinfo {pages} {144}} (\bibinfo {year}
  {1992})}\BibitemShut {NoStop}%
\bibitem [{\citenamefont {G\'{e}ron}(2017)}]{geron2017}%
  \BibitemOpen
  \bibfield  {author} {\bibinfo {author} {\bibfnamefont {A.}~\bibnamefont
  {G\'{e}ron}},\ }\href@noop {} {\emph {\bibinfo {title} {Hands-On Machine
  Learning with Scikit-Learn \& Tensorflow}}}\ (\bibinfo  {publisher} {O'Reilly
  Media, Inc.},\ \bibinfo {year} {2017})\ pp.\ \bibinfo {pages}
  {154--156}\BibitemShut {NoStop}%
\bibitem [{\citenamefont {Sch\H{o}lkopf}\ and\ \citenamefont
  {Smola}(2002)}]{scholkopf2002}%
  \BibitemOpen
  \bibfield  {author} {\bibinfo {author} {\bibfnamefont {B.}~\bibnamefont
  {Sch\H{o}lkopf}}\ and\ \bibinfo {author} {\bibfnamefont {A.}~\bibnamefont
  {Smola}},\ }\href@noop {} {\emph {\bibinfo {title} {Learning with kernels:
  support vector machines, regularization, optimization, and beyond}}}\
  (\bibinfo  {publisher} {MIT Press},\ \bibinfo {year} {2002})\BibitemShut
  {NoStop}%
\bibitem [{\citenamefont {Soydaner}\ and\ \citenamefont
  {Wagemans}(2024)}]{soydaner2023}%
  \BibitemOpen
  \bibfield  {author} {\bibinfo {author} {\bibfnamefont {D.}~\bibnamefont
  {Soydaner}}\ and\ \bibinfo {author} {\bibfnamefont {J.}~\bibnamefont
  {Wagemans}},\ }\href@noop {} {\bibfield  {journal} {\bibinfo  {journal}
  {British Journal of Psychology}\ }\textbf {\bibinfo {volume} {00}},\ \bibinfo
  {pages} {1} (\bibinfo {year} {2024})}\BibitemShut {NoStop}%
\bibitem [{\citenamefont {MacKay}(1998)}]{mackay1998}%
  \BibitemOpen
  \bibfield  {author} {\bibinfo {author} {\bibfnamefont {D.}~\bibnamefont
  {MacKay}},\ }\href@noop {} {\emph {\bibinfo {title} {Introduction to Gaussian
  Processes}}},\ Neural networks and machine learning\ (\bibinfo  {publisher}
  {Springer},\ \bibinfo {year} {1998})\ pp.\ \bibinfo {pages}
  {133--166}\BibitemShut {NoStop}%
\bibitem [{\citenamefont {Alpayd{\i}n}(2014)}]{alpaydin2014}%
  \BibitemOpen
  \bibfield  {author} {\bibinfo {author} {\bibfnamefont {E.}~\bibnamefont
  {Alpayd{\i}n}},\ }\href@noop {} {\emph {\bibinfo {title} {Introduction to
  Machine Learning}}}\ (\bibinfo  {publisher} {MIT Press},\ \bibinfo {year}
  {2014})\ pp.\ \bibinfo {pages} {474--478}\BibitemShut {NoStop}%
\bibitem [{\citenamefont {Rasmussen}\ and\ \citenamefont
  {Williams}(2006)}]{rasmussen2006}%
  \BibitemOpen
  \bibfield  {author} {\bibinfo {author} {\bibfnamefont {C.}~\bibnamefont
  {Rasmussen}}\ and\ \bibinfo {author} {\bibfnamefont {C.}~\bibnamefont
  {Williams}},\ }\href@noop {} {\emph {\bibinfo {title} {Gaussian Processes for
  Machine Learning}}}\ (\bibinfo  {publisher} {MIT Press},\ \bibinfo {year}
  {2006})\BibitemShut {NoStop}%
\bibitem [{\citenamefont {Casten}\ and\ \citenamefont
  {Zamfir}(1996)}]{Casten_1996}%
  \BibitemOpen
  \bibfield  {author} {\bibinfo {author} {\bibfnamefont {R.~F.}\ \bibnamefont
  {Casten}}\ and\ \bibinfo {author} {\bibfnamefont {N.~V.}\ \bibnamefont
  {Zamfir}},\ }\href {https://doi.org/10.1088/0954-3899/22/11/002} {\bibfield
  {journal} {\bibinfo  {journal} {Journal of Physics G: Nuclear and Particle
  Physics}\ }\textbf {\bibinfo {volume} {22}},\ \bibinfo {pages} {1521}
  (\bibinfo {year} {1996})}\BibitemShut {NoStop}%
\bibitem [{\citenamefont {Vogt}\ \emph {et~al.}(2001)\citenamefont {Vogt},
  \citenamefont {Hartmann},\ and\ \citenamefont {Zilges}}]{VOGT2001255}%
  \BibitemOpen
  \bibfield  {author} {\bibinfo {author} {\bibfnamefont {K.}~\bibnamefont
  {Vogt}}, \bibinfo {author} {\bibfnamefont {T.}~\bibnamefont {Hartmann}},\
  and\ \bibinfo {author} {\bibfnamefont {A.}~\bibnamefont {Zilges}},\ }\href
  {https://doi.org/https://doi.org/10.1016/S0370-2693(01)01014-0} {\bibfield
  {journal} {\bibinfo  {journal} {Physics Letters B}\ }\textbf {\bibinfo
  {volume} {517}},\ \bibinfo {pages} {255} (\bibinfo {year}
  {2001})}\BibitemShut {NoStop}%
\bibitem [{\citenamefont {Y\"uksel}\ \emph {et~al.}(2019)\citenamefont
  {Y\"uksel}, \citenamefont {Marketin},\ and\ \citenamefont
  {Paar}}]{PhysRevC.99.034318}%
  \BibitemOpen
  \bibfield  {author} {\bibinfo {author} {\bibfnamefont {E.}~\bibnamefont
  {Y\"uksel}}, \bibinfo {author} {\bibfnamefont {T.}~\bibnamefont {Marketin}},\
  and\ \bibinfo {author} {\bibfnamefont {N.}~\bibnamefont {Paar}},\ }\href
  {https://doi.org/10.1103/PhysRevC.99.034318} {\bibfield  {journal} {\bibinfo
  {journal} {Phys. Rev. C}\ }\textbf {\bibinfo {volume} {99}},\ \bibinfo
  {pages} {034318} (\bibinfo {year} {2019})}\BibitemShut {NoStop}%
\bibitem [{\citenamefont {Nikšić}\ \emph {et~al.}(2014)\citenamefont
  {Nikšić}, \citenamefont {Paar}, \citenamefont {Vretenar},\ and\
  \citenamefont {Ring}}]{NIKSIC20141808}%
  \BibitemOpen
  \bibfield  {author} {\bibinfo {author} {\bibfnamefont {T.}~\bibnamefont
  {Nikšić}}, \bibinfo {author} {\bibfnamefont {N.}~\bibnamefont {Paar}},
  \bibinfo {author} {\bibfnamefont {D.}~\bibnamefont {Vretenar}},\ and\
  \bibinfo {author} {\bibfnamefont {P.}~\bibnamefont {Ring}},\ }\href
  {https://doi.org/https://doi.org/10.1016/j.cpc.2014.02.027} {\bibfield
  {journal} {\bibinfo  {journal} {Computer Physics Communications}\ }\textbf
  {\bibinfo {volume} {185}},\ \bibinfo {pages} {1808} (\bibinfo {year}
  {2014})}\BibitemShut {NoStop}%
\bibitem [{\citenamefont {Garvey}\ \emph {et~al.}(1969)\citenamefont {Garvey},
  \citenamefont {Gerace}, \citenamefont {Jaffe}, \citenamefont {Talmi},\ and\
  \citenamefont {Kelson}}]{RevModPhys.41.S1}%
  \BibitemOpen
  \bibfield  {author} {\bibinfo {author} {\bibfnamefont {G.~T.}\ \bibnamefont
  {Garvey}}, \bibinfo {author} {\bibfnamefont {W.~J.}\ \bibnamefont {Gerace}},
  \bibinfo {author} {\bibfnamefont {R.~L.}\ \bibnamefont {Jaffe}}, \bibinfo
  {author} {\bibfnamefont {I.}~\bibnamefont {Talmi}},\ and\ \bibinfo {author}
  {\bibfnamefont {I.}~\bibnamefont {Kelson}},\ }\href
  {https://doi.org/10.1103/RevModPhys.41.S1} {\bibfield  {journal} {\bibinfo
  {journal} {Rev. Mod. Phys.}\ }\textbf {\bibinfo {volume} {41}},\ \bibinfo
  {pages} {S1} (\bibinfo {year} {1969})}\BibitemShut {NoStop}%
\bibitem [{\citenamefont {Barea}\ \emph {et~al.}(2008)\citenamefont {Barea},
  \citenamefont {Frank}, \citenamefont {Hirsch}, \citenamefont {Isacker},
  \citenamefont {Pittel},\ and\ \citenamefont
  {Vel\'azquez}}]{PhysRevC.77.041304}%
  \BibitemOpen
  \bibfield  {author} {\bibinfo {author} {\bibfnamefont {J.}~\bibnamefont
  {Barea}}, \bibinfo {author} {\bibfnamefont {A.}~\bibnamefont {Frank}},
  \bibinfo {author} {\bibfnamefont {J.~G.}\ \bibnamefont {Hirsch}}, \bibinfo
  {author} {\bibfnamefont {P.~V.}\ \bibnamefont {Isacker}}, \bibinfo {author}
  {\bibfnamefont {S.}~\bibnamefont {Pittel}},\ and\ \bibinfo {author}
  {\bibfnamefont {V.}~\bibnamefont {Vel\'azquez}},\ }\href
  {https://doi.org/10.1103/PhysRevC.77.041304} {\bibfield  {journal} {\bibinfo
  {journal} {Phys. Rev. C}\ }\textbf {\bibinfo {volume} {77}},\ \bibinfo
  {pages} {041304} (\bibinfo {year} {2008})}\BibitemShut {NoStop}%
\bibitem [{\citenamefont {Lundberg}\ and\ \citenamefont
  {Lee}(2017)}]{lundberg2017}%
  \BibitemOpen
  \bibfield  {author} {\bibinfo {author} {\bibfnamefont {S.}~\bibnamefont
  {Lundberg}}\ and\ \bibinfo {author} {\bibfnamefont {S.}~\bibnamefont {Lee}},\
  }\href
  {https://proceedings.neurips.cc/paper_files/paper/2017/file/8a20a8621978632d76c43dfd28b67767-Paper.pdf}
  {\bibfield  {journal} {\bibinfo  {journal} {Advances in Neural Information
  Processing Systems}\ }\textbf {\bibinfo {volume} {30}},\ \bibinfo {pages}
  {061302} (\bibinfo {year} {2017})}\BibitemShut {NoStop}%
\bibitem [{\citenamefont {Shapley}(1953)}]{shapley1953}%
  \BibitemOpen
  \bibfield  {author} {\bibinfo {author} {\bibfnamefont {L.}~\bibnamefont
  {Shapley}},\ }\href@noop {} {\bibfield  {journal} {\bibinfo  {journal}
  {Contributions to the Theory of Games II}\ ,\ \bibinfo {pages} {307–317}}
  (\bibinfo {year} {1953})}\BibitemShut {NoStop}%
\bibitem [{\citenamefont {Winter}(2002)}]{winter2002}%
  \BibitemOpen
  \bibfield  {author} {\bibinfo {author} {\bibfnamefont {E.}~\bibnamefont
  {Winter}},\ }\href@noop {} {\bibfield  {journal} {\bibinfo  {journal}
  {Handbook of Game Theory with Economic Applications}\ }\textbf {\bibinfo
  {volume} {3}},\ \bibinfo {pages} {2025–2054} (\bibinfo {year}
  {2002})}\BibitemShut {NoStop}%
\bibitem [{Note1()}]{Note1}%
  \BibitemOpen
  \bibinfo {note}
  {Https://shap-lrjball.readthedocs.io/en/latest/examples.html}\BibitemShut
  {NoStop}%
\end{thebibliography}%

\end{document}